\documentclass[journal]{IEEEtran}

\usepackage{graphicx,amssymb,lineno}
\usepackage{amsmath,amsfonts,amssymb}

\usepackage{algorithm}
\usepackage{algorithmic}
\usepackage{cite}
\usepackage[usenames]{color}

\usepackage{multirow}

\usepackage{graphicx,graphics,color,epsfig,subfigure,graphpap,rotate}
\usepackage{times, verbatim, subfigure, epsfig, graphicx, latexsym, amsmath}
\usepackage{url}
\usepackage{subfigure}
\begin{document}
%
\title{Boosting Spatial Reuse via Multiple Paths Multi-Hop Scheduling for Directional mmWave WPANs}



\author{Yong~Niu,
        Chuhan~Gao,
        Yong~Li,~\IEEEmembership{Member,~IEEE,}
        Depeng~Jin,
        Li~Su,
        and Dapeng (Oliver) Wu,~\IEEEmembership{Fellow,~IEEE} %
        
\thanks{Copyright (c) 2015 IEEE. Personal use of this material is permitted. However, permission to use this material for any other purposes must be obtained from the IEEE by sending a request to pubs-permissions@ieee.org.}
\thanks{Y. Niu, Chuhan Gao, Y. Li, D. Jin, L. Su are with State Key Laboratory on
 Microwave and Digital Communications, Tsinghua National Laboratory for Information
 Science and Technology (TNLIST), Department of Electronic Engineering, Tsinghua
 University, Beijing 100084, China (E-mails: liyong07@tsinghua.edu.cn).} 
\thanks{
 D. O. Wu is with the Department of Electrical and Computer Engineering,
University of Florida, Gainesville, FL 32611-6130, USA.} %
\thanks{This work was partially supported by the National Natural Science
Foundation of China (NSFC) under grant No. 61201189 and 61132002, National High Tech (863) Projects
under Grant No. 2011AA010202, Research Fund of Tsinghua University under No. 2011Z05117 and
20121087985, and Shenzhen Strategic Emerging Industry Development Special Funds under No.
CXZZ20120616141708264.}
}%

\maketitle

\begin{abstract}

With huge unlicensed bandwidth available in most parts of the world, millimeter wave (mmWave)
communications in the 60 GHz band has been considered as one of the most promising candidates to
support multi-gigabit wireless services. Due to high propagation loss of mmWave channels,
beamforming is likely to become adopted as an essential technique. Consequently, transmission in 60 GHz band is
inherently directional. Directivity enables concurrent transmissions (spatial reuse), which can
be fully exploited to improve network capacity. In this paper, we propose a multiple paths
multi-hop scheduling scheme, termed MPMH, for mmWave wireless personal area networks, where the
traffic across links of low channel quality is transmitted through multiple paths of multiple
hops to unleash the potential of spatial reuse. We formulate the problem of multiple paths
multi-hop scheduling as a mixed integer linear program (MILP), which is generally NP-hard. To
enable the implementation of the multiple paths multi-hop transmission in practice, we propose a
heuristic scheme including path selection, traffic distribution, and multiple paths multi-hop
scheduling to efficiently solve the formulated problem.   
Finally, through extensive simulations, we demonstrate MPMH achieves near-optimal network
performance in terms of transmission delay and throughput, and enhances the network performance
significantly compared with existing protocols.

\end{abstract}

\section{Introduction}\label{S1}

Recently, millimeter wave (mmWave) communications in the 60 GHz band has attracted considerable
interest from academia, industry, and standards bodies. Due to its huge unlicensed bandwidth (i.e.,
up to 7 GHz in the USA), 60 GHz communications is able to support multi-gigabit wireless services,
such as high-speed data transfer between devices (e.g., cameras, pads, and personal computers),
wireless gigabit ethernet, wireless gaming, and real-time streaming of both the compressed and
uncompressed high definition television. Meanwhile, rapid progress in 60-GHz
mm-wave circuits, including on-chip and in-package antennas, radio frequency
(RF) power amplifiers (PAs), low-noise amplifiers
(LNAs), voltage-controlled oscillators (VCOs), mixers, and
analog-to-digital converters (ADCs) has accelerated popularization of wireless products and services in the 60 GHz band \cite{CMOS,CMOS2,CMOS3}. In terms of
standardization, several standards have been defined for indoor wireless personal area networks
(WPAN) and wireless local area networks (WLAN), for example, ECMA-387 \cite{ECMA_387} , IEEE
802.15.3c \cite{IEEE_802.15.3c}, and IEEE 802.11ad \cite{IEEE_802.11ad}. Due to high carrier
frequency, mmWave communication systems are fundamentally different from other existing communication
systems using lower carrier frequencies (e.g., from 900 MHz to 5 GHz). Therefore, network
architectures and protocols, especially medium access control (MAC) mechanism, require new thinking
to fully unleash the potential of mmWave communications.

MmWave communications in the 60 GHz band suffers from high propagation loss. The free space
propagation loss scales as the square of the wavelength, which indicates that the free space
propagation loss at 60 GHz band is 28 decibels (dB) more than that at 2.4 GHz \cite{singh_outdoor}.
Thus, 60 GHz communications is mainly used for short-range indoor communications. To combat severe
channel attenuation, high gain directional antennas are utilized at both the transmitter and
receiver, where beamforming has been adopted as an essential technique to search for the best
antenna weight vectors (AWV) \cite{beam_training, Beamtraining2, Beam_xiao}.

With directional transmission, the third party nodes cannot perform carrier sense as in IEEE 802.11 to avoid contention with current transmission, which is know as the deafness problem \cite{mao}. On the other hand, there is less interference between links, and thus concurrent
transmission (spatial reuse) can be exploited to greatly increase the network capacity. However,
on one hand, for flows across links of low channel quality or with significantly higher traffic demand, more network resources (e.g., time
slots) are needed. In this case, these extra resources cannot be utilized efficiently due to less
spatial reuse. Consequently, system performance is degraded significantly. Thus, more efficient
transmission mechanisms are needed to address the performance degradation when there are flows across links of low channel quality or with significantly higher traffic demand. For example, in \cite{Qiao_7}, a
traffic flow of a long hop is transmitted over multiple short hops to improve flow throughput.
Similarly, in D-CoopMAC \cite{chenqian}, the direct long link is replaced by a two-hop link if a
relay that has higher rate links with both source and destination exists. On the other hand, for
flows supporting applications of high throughput, such as high-definition television (HDTV), much more time slots are needed to
satisfy their throughput requirements in TDMA-based protocols such as IEEE 802.15.3c. Consequently,
there are not enough time slots allocated for other flows, which may lead to serious unfairness in
WPANs. Besides, the number of flows scheduled each time in these protocols will be very small. Therefore, fully exploiting spatial reuse to improve the throughput of these flows while achieving
high system performance is an important and challenging issue.


Recently, the hybrid beamforming structure has been proposed to obtain the multiplexing gain of Multiple-Input Multiple-Output (MIMO) and also provide high beamforming gain to overcome high propagation loss in mmWave bands \cite{mmW_mimo}. In the hybrid beamforming structure, multiple RF links can be established via the design of the digital precoder and the analog beamformer to reap the spatial multiplexing gain of MIMO \cite{mmW_mimo2, survey_own}. To improve the throughput of flows of low channel quality or with high traffic demand, transmitting these flows through multiple paths concurrently similar to the multiple RF links in the hybrid beamforming structure will be an effective way, and should be considered in the design of scheduling schemes for mmWave communications.


In this paper, motivated by the spatial multiplexing structure in hybrid beamforming, we propose a novel multiple paths multi-hop scheduling scheme (MPMH) to boost the
spatial reuse in mmWave WPANs. By transmitting the traffic of flows with low channel quality on the direct paths or
high traffic demand through multiple multi-hop paths, the potential of spatial reuse is unleashed,
and concurrent transmissions on these paths are fully exploited to enhance the system performance
in terms of flow and network throughput. The contributions of this paper are three-fold, which are
summarized as follows.


\begin{itemize}
\item We formulate the multiple paths multi-hop scheduling problem into a mixed integer
linear program (MILP), i.e., to minimize the number of time slots by multiple paths multi-hop
transmissions with the traffic demand of all flows accommodated. Concurrent transmissions, i.e.,
spatial reuse, are explicitly considered under the signal to interference plus noise ratio (SINR) interference model in this formulated
problem.
\item We propose an efficient and practical scheme to solve the formulated NP-hard problem by three heuristic algorithms of path
selection, traffic distribution, and transmission scheduling. In the transmission scheduling
algorithm, concurrent transmissions are enabled if the SINR of each link is able to support its
transmission rate.
\item Extensive
simulations are carried out to demonstrate the near-optimal network performance of MPMH, and
superior performance in terms of delay and throughput compared with other existing protocols. Performance under different maximum number of hops on paths is also investigated to provide guidelines for the choice of this parameter in practice.

\end{itemize}

The rest of this paper is organized as follows. We discuss related work
on concurrent transmission scheduling for mmWave WPANs in section \ref{S7}. In section \ref{S3}, we present the system model
and illustrate our basic idea by an example. We formulate the optimal multiple paths multi-hop
scheduling problem as an MILP in section \ref{S4}. The MPMH scheme is illustrated in section
\ref{S5}. In section \ref{S6}, we evaluate the performance of MPMH under various traffic patterns and simulation parameters. Finally, we conclude
this paper in section \ref{S8}.


\section{Related Work}\label{S7}

Transmission scheduling for mmWave communications in the 60 GHz band has been investigated in the
literature \cite{Qiao_7, Qiao, mao, Qiao_6, Qiao_15, EX_Region, chen_2, Spatial_reuse_TDMA,mao_12,
mao_13,chenqian,MRDMAC,Gong}. Since ECMA-387 \cite{ECMA_387} and IEEE 802.15.3c
\cite{IEEE_802.15.3c} adopted time division multiple access (TDMA), some work is also based on TDMA \cite{EX_Region, mao_12, mao_13, Qiao_6,
Qiao_15, Qiao}. In two protocols based on IEEE 802.15.3c, multiple links are scheduled to
communicate in the same slot if the multi-user interference (MUI) is below a specific threshold
\cite{Qiao_6,Qiao_15}. Cai \emph{et al.} \cite{EX_Region} introduced the concept of exclusive
region (ER) to enable concurrent transmissions, and derived the ER conditions that concurrent
transmissions always outperform TDMA. Qiao \emph{et al.} \cite{Qiao} proposed a concurrent
transmission scheduling algorithm for an indoor IEEE 802.15.3c WPAN, where non-interfering and
interfering links are scheduled to transmit concurrently to maximize the number of flows with the
quality of service requirement of each flow satisfied. Furthermore, multi-hop concurrent
transmissions (MHCT) are enabled to address the link outage problem (blockage) and to combat huge path
loss to improve flow throughput \cite{Qiao_7}. MHCT, however, only transmits traffic of flows
through multiple hops on one path, and does not exploit the concurrent transmissions on multiple
paths to improve flow throughput and network throughput. In IEEE 802.15.3c, during the random
access period, the piconet controller is in the omni-directional mode to solve the deafness
problem, which may not be feasible for mmWave systems of the multi-gigabit domain with highly
directional transmissions. For bursty traffic patterns such as the Interrupted Poisson Process (IPP), TDMA based protocols may result in over-allocated
medium access time for some users while resulting in under-allocated medium access time for others.

There is some other work based on a central controller to coordinate the transmissions in WPANs
\cite{Gong,mao,MRDMAC,chen_2,chenqian}. Gong \emph{et al.} \cite{Gong} proposed a directional carrier sense multiple access with collision avoidance (CSMA/CA) protocol, which mainly focuses on solving the deafness problem. It exploits virtual carrier
sensing, and depends on the piconet coordinator (PNC) to distribute the network allocation vector
(NAV) information. However, it does not exploit the spatial reuse fully and also does not consider
multiple paths multi-hop transmissions. In the multi-hop relay directional MAC protocol (MRDMAC), if a
wireless terminal (WT) is lost, the access point (AP) will choose a WT among the live WTs to act as
a relay to the lost node \cite{MRDMAC}. However, MRDMAC does not fully exploit spatial reuse since
most transmissions go through the piconet coordinator (PNC). Chen \emph{et al.} \cite{chen_2}
proposed a spatial reuse strategy to schedule two different service periods (SPs) to overlap with each other for an
IEEE 802.11 ad WPAN. Recently, Son \emph{et al.}
\cite{mao} proposed a frame based directional MAC protocol (FDMAC).
The high efficiency of FDMAC is achieved by amortizing the
scheduling overhead over multiple concurrent transmissions in a row.
The core of FDMAC is the Greedy Coloring algorithm, which fully
exploits spatial reuse and greatly improves the network throughput
compared with MRDMAC \cite{MRDMAC} and memory-guided directional MAC
(MDMAC) \cite{MDMAC}. FDMAC also has a good fairness performance
and low complexity. FDMAC, however, does not consider multiple paths multi-hop
transmission to improve flow throughput of links with poor quality. Chen \emph{et al.} \cite{chenqian} proposed a directional
cooperative MAC protocol, termed D-CoopMAC, to coordinate the uplink channel access among stations
in an IEEE 802.11ad WLAN. In D-CoopMAC, the multirate capability of links is exploited to select a
relay station for the direct link; when the two-hop link outperforms the direct link, the latter
will be replaced by the former. In D-CoopMAC, however, spatial reuse is not considered since most
transmissions go through the access point (AP). Most of the above work ignores the negative effect
of links with poor channel quality on network performance, such as delay, throughput, and the
number of flows scheduled each time. The work above does not boost the potential of spatial
reuse via multiple paths multi-hop transmissions.


There is also some related work on maximum independent set and protocol model based scheduling. For
protocol model based scheduling, interference is modeled as an interference graph, where each
vertex represents a link in the wireless network \cite{Xu_11, Xu_12, Xu_18}. Two vertices form an
edge if these two links cannot be scheduled for concurrent transmissions. This simple interference
model does not take the unique features of mmWave links, e.g., directivity, into consideration. Xu
\emph{et al.} \cite{Xu_mis} proposed a constant approximation algorithm for one-slot link
scheduling in arbitrary networks under the SINR interference model, where a transmission is
successful if the received SINR is more than some threshold. However, it does not consider the
directivity of mmWave links, and the interference is only related to the distance between nodes in
a two-dimensional Euclidean space. This lack of consideration is unreasonable for mmWave WPANs, especially since the SINR
thresholds for all links were kept the same. To the best of our knowledge, we are the first to exploit multiple paths multi-hop transmissions for
boosting the spatial reuse gain under the SINR interference model.


\section{System Overview}\label{S3}

\subsection{System Model}\label{S3-a}

We consider a typical indoor WPAN system composed of several wireless nodes (WNs) and a single
piconet controller (PNC) \cite{IEEE_802.15.3c}. The PNC synchronizes the clocks of other nodes and
schedules the medium access of all the nodes to accommodate their traffic demands. The WNs and PNC
have electronically steerable directional antennas to support directional transmissions between WNs
or between the WN and the PNC by beamforming. The system is partitioned into non-overlapping time slots of
equal length, and runs a bootstrapping program \cite{bootstrapping}, by which each device knows the
up-to-date network topology and the location information of other devices. With this
information, the beamforming between nodes can be completed in a short time, and each node can direct its antenna towards other nodes.

We assume there are $V$ flows with traffic transmission demand in the network. For flow $v$, we
denote its traffic demand as $d_v$, and numerically $d_v$ is equal to the number of packets to be
transmitted. For 60 GHz wireless channels, non-line-of-sight (NLOS) transmissions suffer from
higher attenuation than line-of-sight (LOS) transmissions \cite{NLOS,NLOS2}. In Ref. \cite{NLOS}, the path loss exponent in the LOS hall is 2.17, while the path loss exponent in the NLOS hall is 3.01. Thus, if the distance between the transmitter and the receiver is 10 m, the gap between the path loss of LOS hall and NLOS hall is about 10 dB. In a power-limited regime, a 10 dB power loss requires a 10-fold reduction of transmission rate to maintain the same reliability. On the other hand, NLOS transmission in the 60 GHz band also suffers from a shortage of multipath, and restricting to the LOS path can maximize the power efficiency since the LOS path is strongest \cite{MRDMAC}. To achieve high data rate
transmission and maximize the power efficiency, we consider the directional LOS
transmission case in this paper. For each directional link $i$, we denote its sender and receiver
as $s_i$ and $r_i$ respectively. Then according to the path loss model \cite{Qiao}, the received
signal power, denoted by $P_r$ (mW), can be calculated as
\begin{equation}
{P_r} = {k_0}{P_t}{l_{s_ir_i}^{ - \gamma }},
\end{equation}
where ${{P_t}}$ (mW) is the transmission power, $k_0 = {10^{PL({d_0})/10}}$ is the constant scaling factor
corresponding to the reference path loss ${PL({d_0})}$ (dB) with $d_0$ equal to 1 m, ${{l_{s_ir_i}}}$ (m) is
the distance between node $s_i$ and node $r_i$, and $\gamma $ is the path loss exponent
\cite{Qiao}. In this paper, we assume fixed transmission power.

We denote the transmission rate of the direct link of flow $v$ as $c_v$, and numerically $c_v$ is equal to the number
of packets the link can transmit in one time slot. The transmission rates of links are obtained by a channel transmission rate
measurement procedure \cite{tvt_own}. In this procedure, the sender of each flow transmits
measurement packets to the receiver first. After measuring the signal to noise ratio (SNR) of these
packets, the receiver obtains the achievable transmission rate, and appropriate modulation and
coding scheme (MCS) by the correspondence table about the SNR and MCS. Under low
user mobility, the procedure is usually executed, and the transmission rates of links are updated periodically.

Directional transmissions enable less interference between links, and concurrent transmissions of links can be supported to improve network capacity. Due to the limited range for mmWave WPANs, the interference between links cannot be neglected \cite{EX_Region}. Thus, we adopt the SINR model for concurrent transmission \cite{Qiao, Xu_mis}. For link $i$ and link $j$, the received power from $s_i$ to $r_j$ can be
calculated as
\begin{equation}
P_r^{i,j} = {f_{s_i,r_j}}k_0{P_t}{l^{ - \gamma }_{s_ir_j}},
\end{equation}
where $f_{s_i,r_j}$ indicates whether $s_i$ and $r_j$ direct their beams towards each other. If $s_i$ and $r_j$ direct their beams towards each other,
${{f_{s_i,r_j}}}=1$; otherwise, ${{f_{s_i,r_j}}}=0$. Then, the received SINR at $r_j$, denoted by $SINR_j$,
can be calculated as
\begin{equation}
SIN{R_{j}} = \frac{{k_0{P_t}{l^{ - \gamma }_{s_jr_j}}}}{{W{N_0} + \rho \sum\limits_{i \ne j}
{{f_{s_i,r_j}}k_0{P_t}{l^{ - \gamma }_{s_ir_j}}} } },
\end{equation}
where $\rho$ is the multi-user interference (MUI) factor related to the cross correlation of signals from different links,
$W$ (Hz) is the bandwidth, and ${{N_0}}$ (mW/Hz) is the one-sided
power spectra density of white Gaussian noise \cite{Qiao}. For each link $j$, we denote the minimum SINR to support its transmission rate
${c_{j}}$ as $MS({c_{j}})$. Therefore, concurrent transmissions are supported if the SINR of each
link $j$ is larger than or equal to $MS({c_{j}})$.

\begin{figure*}[htbp]
\begin{minipage}[t]{0.5\linewidth}
\centering
\includegraphics[width=8.5cm]{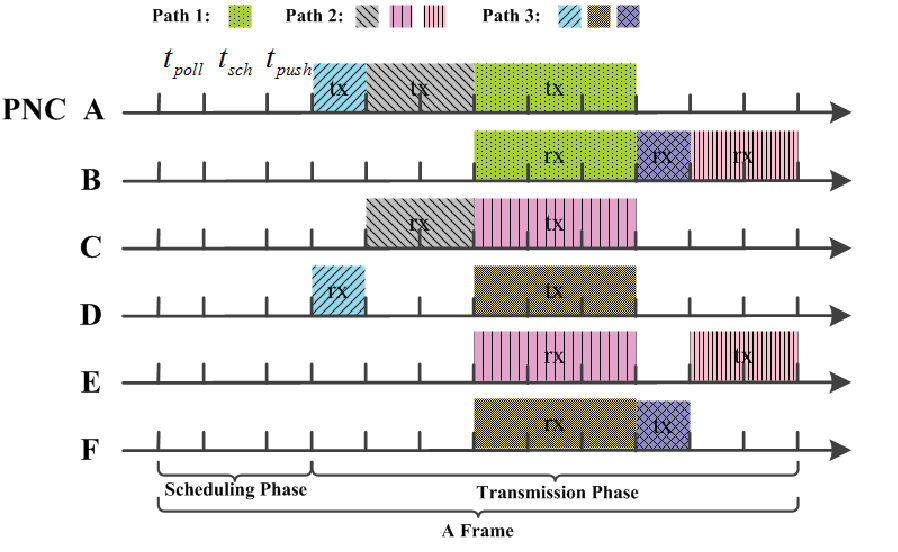}
\centerline{\small (a) A MPMH frame}
\end{minipage}%
\begin{minipage}[t]{0.5\linewidth}
\centering
\includegraphics[width=7.5cm]{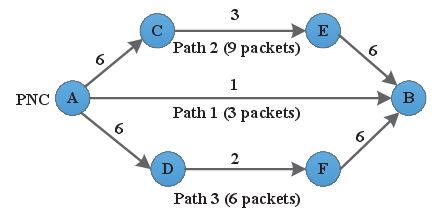}
\centerline{\small (b) Transmission path illustration of MPMH}
\end{minipage}
\caption{An example of MPMH operation in a WPAN of six nodes.}
\label{fig:mpmh operation} 
\vspace*{-3mm}
\end{figure*}

In MPMH, we adopt a frame based scheduling scheme, which is shown in Fig. \ref{fig:mpmh operation}(a).
Node A is the PNC, and other nodes are WNs. In MPMH, time is divided into a sequence of
non-overlapping frames \cite{mao}. Each frame consists of a scheduling phase and a transmission
phase. In the scheduling phase, after all of the WNs steer their antennas towards the PNC, the PNC
polls the WNs successively for their traffic demand in time ${t_{poll}}$. Then the PNC
computes a schedule to accommodate the traffic demand in time ${t_{sch}}$. Finally, the
PNC pushes the schedule to WNs in time ${t_{push}}$. In the transmission phase, all
devices communicate with each other following the schedule until their traffic demand is cleared.
Spatial reuse gain can be boosted via multiple paths multi-hop transmission in the transmission
phase to improve flow as well as network throughput, which depends on the scheduling solution.



\subsection{Problem Overview} \label{S3-b}


For flows with low channel quality on their direct paths or high traffic demand, to improve throughput, their traffic
should be transmitted through multiple paths to unleash the potential of spatial reuse of hops on
these paths. Meanwhile, the traffic demand of flows should be accommodated with a minimum number of
time slots in the transmission phase to maximize the transmission efficiency.

Now, we present an example to illustrate the operation of MPMH and our basic idea. We assume a WPAN
of six devices in the network, denoted as $A$, $B$, $C$, $D$, $E$, and $F$, as shown in Fig.
\ref{fig:mpmh operation}(b). The numbers above the directional edges between nodes represent their
transmission rates, and numerically are equal to the numbers of packets these links can transmit in
one time slot. If there are 18 packets to be transmitted from $A$ to $B$, then with a transmission
rate of 1 packet per time slot, 18 slots are needed in the transmission phase to clear this traffic
demand. By MPMH, we select three paths from $A$ to $B$, i.e., the direct path from $A \to B$ (path
1), the path of $A \to C \to E \to B$ (path 2), and the path of $A \to D \to F \to B$ (path 3).
Then we distribute 3 packets to path 1, 9 packets to path 2, and 6 packets to path 3. Afterwards,
MPMH computes a schedule as illustrated in Fig. \ref{fig:mpmh operation}(a), where each colored box represents a time slot. The schedule has 5
pairings, and in the third pairing of the schedule, links $A \to B$, $C \to E$, and $D \to F$
transmit concurrently for 3 time slots. This schedule clears the packets from $A$ to $B$ in 9 time
slots, and by transmitting this flow on three paths, concurrent transmissions of the three links in
the third pairing are enabled to improve the efficiency of transmission significantly.


\section{Problem Formulation}\label{S4}

In this section, we formulate the problem of optimal multiple paths multi-hop
scheduling into a mixed integer linear program (MILP) based on the problem formulation in FDMAC \cite{mao}.

\subsection{Problem Formulation and Analysis}\label{S4-a}

From traffic demand polling, we assume there are $V$ flows to be scheduled by the PNC. For flows across links with a low transmission rate or with a high traffic demand, more time
slots are needed to accommodate their traffic demand. Consequently, these extra time slots cannot be used for
spatial reuse, which degrades system performance significantly. Thus, flows with lower transmission rate on the direct paths or with higher traffic demand have higher priority to be transmitted through multiple paths for better utilization of spatial reuse.
In the following, we propose a criterion to decide whether a flow needs to be transmitted through multiple paths.

For each flow $v$, we define its traffic demand intensity as arrived traffic demand averaged over a relatively long time, which is denoted by $\overline {{D_v}}$. The transmission rate of its direct path is denoted by $c_v$. Then if flow $v$ satisfies
\begin{equation}
\frac{{ {c_v}/\overline {{D_v}}}}{{\mathop {{\rm{avg}}}\limits_{u \in V} ({c_u}/\overline {{D_u}})}} < \varepsilon,\label{flow_requirement}
\end{equation}
then traffic of flow $v$ will be transmitted through multiple paths. $\varepsilon $ is defined to
control the number of flows transmitted through multiple paths. With a larger $\varepsilon $, there
may be more flows transmitted through multiple paths; otherwise, fewer flows will be transmitted
through multiple paths. Specially, if $c_v$ is equal to 0, i.e., the direct path of flow $v$ is blocked, flow $v$ will be transmitted through multiple paths.

For the fairness among flows, if there is no multi-path multi-hop transmission, flows across direct paths of low channel quality or with high traffic demand will occupy a large number of time slots exclusively. With multi-path multi-hop transmission enabled, more flows are able to share the time resources more fully by the concurrent transmissions (spatial reuse). Thus, the fairness performance without multi-path multi-hop transmission will be worse than that with multi-path multi-hop transmission. Therefore, our scheme improves the fairness among flows by the multi-path multi-hop transmission.

 We denote the number of paths of the $v^{th}$ flow
as $M_v$. We also denote the number of hops of the $p^{th}$ path of flow $v$ as $H_{vp}$. For flow $v$,
the traffic demand distributed to the $p^{th}$ path is denoted as $d_{vp}$. We denote the $i^{th}$ hop
link of the $p^{th}$ path of flow $v$ as $(v,p,i)$, and its transmission rate by $c_{vpi}$. For link $(v,p,i)$ and $(u,q,j)$, we define an
indicator variable $I_{vpi,uqj}$. $I_{vpi,uqj}$ is equal to 1 if $(v,p,i)$ and $(u,q,j)$ are
adjacent, i.e., have common vertices; otherwise, $I_{vpi,uqj}$ is equal to 0.

If there are $K$ pairings in the schedule to accommodate the traffic demand of flows, we denote the
number of time slots of the ${k^{th}}$ pairing as $\delta^{k}$ \cite{mao}. We also define $a_{vpi}^k$ to indicate
whether link $(v,p,i)$ is scheduled in the ${k^{th}}$ pairing. If link $(v,p,i)$ is scheduled in the ${k^{th}}$ pairing, $a_{vpi}^k$ is equal to 1;
otherwise, $a_{vpi}^k$ is equal to 0. To optimize system performance, the traffic demand of flows should be
cleared in the shortest time \cite{mao}. Thus, the problem of optimal multiple paths multi-hop scheduling (P1) is
formulated as follows.

\begin{equation}
 \min \sum\limits_{k = 1}^K
{{\delta ^k}} \label{OBJ} \hspace{7.0cm}
\end{equation}
\hspace{0.15cm}s. t.

\begin{equation}
\begin{array}{l}\hspace{-0.85cm}
\sum\limits_{k = 1}^K {a_{vpi}^k} \left\{ {\begin{array}{*{20}{c}}
{ = 1,\;{\rm{if}}\;{d_{vp}} > 0\;\&\;i \le {H_{vp}}},\\
{ = 0,\;{\rm{otherwise}};\hspace{1.65cm}}
\end{array}} \right.
\hspace{0.3cm}\forall \;v,p,i \label{CONS1}
\end{array}
\end{equation}

\begin{equation}
\begin{array}{l}\hspace{-0.08cm}
a_{vpi}^k \in \left\{ {\begin{array}{*{20}{c}}
{\{ 0,1\} ,\;{\rm{if}}\;{d_{vp}} > 0\;\&\;i \le {H_{vp}}},\\
{\{ 0\},\;{\rm{otherwise}};}\hspace{2cm}
\end{array}} \right.
\hspace{0.3cm}\forall\;v,p,i,k\label{CONS2}
\end{array}
\end{equation}

\begin{equation}
\begin{array}{l}\hspace{-0.0cm}
\sum\limits_{k = 1}^K {({\delta ^k} \cdot a_{vpi}^k)}\left\{ {\begin{array}{*{20}{c}}{\hspace{-0.2cm} \ge
\left\lceil {\frac{{d_{vp}}}{{c_{vpi}}} } \right\rceil ,\;{\rm{if}}\;{d_{vp}} >
0\;\&\;i \le {H_{vp}}},\\
{\hspace{-0.2cm}=0,\hspace{0.9cm}{\rm{otherwise}};}\hspace{1.7cm}
\end{array}\;} \right.
\hspace{-0.2cm}\forall\;v,p,i\label{CONS3}
\end{array}
\end{equation}

\begin{equation}
\begin{array}{l}\hspace{-4.4cm}
\sum\limits_{p = 1}^{{M_v}} {{d_{vp}}}  = {d_v};\;\; \forall\;v\label{CONS4}
\end{array}
\end{equation}

\begin{equation}
\begin{array}{l}\hspace{-4.4cm}
\sum\limits_{i = 1}^{{H_{vp}}} {a_{vpi}^k \le 1}; \;\;\forall\;v,p,k\;\label{CONS5}
\end{array}
\end{equation}

\begin{equation}
\begin{array}{l}\hspace{-0.4cm}
\hspace{0.1cm}a_{vpi}^k + a_{uqj}^k \le 1,\;\;{\rm{if}}\;\;{I_{vpi,uqj}} = 1;\\
\hspace{2cm}{\rm{for}}\;{\rm{all}}\;k,\;{\rm{any}}\;{\rm{two}}\;{\rm{links}}\;(v,p,i),\left(
{u,q,j} \right)\label{CONS6}
\end{array}
\end{equation}

\begin{equation}
\begin{array}{l}\hspace{-0.2cm}
\hspace{0.0cm}\sum\limits_{k = 1}^{K^*} {a_{vpi}^k}  \ge \sum\limits_{k = 1}^{K^*} {a_{vp(i + 1)}^k},\;\;{\rm{if}}\; {H_{vp}} > 1;\;\\
\;\;\;\;\;\;\;\;\;\;\;\;\;\hspace{1.0cm}\forall\;v,p,i = 1 \thicksim ({H_{vp}} -
1), \;{K^*} =1 \thicksim K\label{CONS7}
\end{array}
\end{equation}


\begin{equation}
\begin{array}{l}\hspace{-0.7cm}
\frac{{{k_0}{P_t}{l^{ - \gamma }_{{s_{vpi}},\;{r_{vpi}}}}a_{vpi}^k}}{{W{N_0} + \rho \sum\limits_{u = 1}^V {\sum\limits_{q = 1}^{{M_u}} {\sum\limits_{j = 1}^{{H_{uq}}} {{f_{{s_{uqj}},\;{r_{vpi}}}}a_{uqj}^k{k_0}{P_t}{l^{ - \gamma }_{{s_{uqj}},\;{r_{vpi}}}}} } } }} \\\hspace{1.5cm}\ge MS({c_{vpi}})
 \times a_{vpi}^k.\hspace{0.5cm}\forall \;v,p,i,k\;
\end{array}\label{CONS8}
\end{equation}

These constraints are explained as follows.

\begin{itemize}

\item Constraint (\ref{CONS1}) indicates for each link
$(v,p,i)$, if there is traffic distributed to it, then it should be scheduled once in one pairing
of the frame.

\item Constraint (\ref{CONS2}) indicates for each link
$(v,p,i)$, if there is traffic distributed to it, then $a_{vpi}^k$ is a binary variable; otherwise,
$a_{vpi}^k$ is equal to 0.

\item Constraint (\ref{CONS3}) indicates for each link
$(v,p,i)$, if there is traffic distributed to it, then this traffic should be accommodated in the
frame.

\item Constraint (\ref{CONS4}) indicates for flow $v$, the sum
of traffic distributed in all paths should be equal to its traffic $d_v$.

\item Constraint (\ref{CONS5}) indicates links in the same
path cannot be scheduled in the same pairing due to the inherent order of transmission in each
path. Preceding hops should be scheduled ahead of hops behind on the path since nodes behind on the path are able to relay the packets after receiving the packets from preceding nodes.

\item Constraint (\ref{CONS6}) indicates due to the half-duplex
assumption, adjacent links cannot be scheduled in the same pairing.

\item Constraint (\ref{CONS7}) indicates due to the inherent
order of transmission in each path, the $i^{th}$ hop link of the $p^{th}$ path of flow $v$ should
be scheduled ahead of the ${(i+1)}^{th}$ hop link. Constraint (\ref{CONS7}) represents a group of
constraints since $K^*$ varies from 1 to $K$.

\item Constraint (\ref{CONS8}) indicates to enable concurrent
transmissions, the SINR of each link in the same pairing should be able to support its transmission
rate.

\end{itemize}

\subsection{Problem Reformulation}

We can observe that in the constraints (\ref{CONS1}), (\ref{CONS2}), and (\ref{CONS3}) of problem
(P1), the condition ``${\rm{if}}\;{d_{vp}}
> 0$'' has the variable $d_{vp}$. This problem form is intractable. If we select the paths for each flow that needs multiple paths multi-hop
transmission by a heuristic path selection algorithm, these paths will have traffic, and their
conditions ``${\rm{if}}\;{d_{vp}}
> 0$'' will be met. In this case, we can remove ``${\rm{if}}\;{d_{vp}}
> 0$'' from constraints (\ref{CONS1}), (\ref{CONS2}), and (\ref{CONS3}), and problem (P1)
becomes a mixed integer nonlinear program (MINLP). However, it is still difficult to obtain the
optimal solution since constraints (\ref{CONS3}) and (\ref{CONS8}) are nonlinear. If the nonlinear
terms of problem (P1) can be linearized, problem (P1) will become a standard mixed integer linear
program (MILP), which can be solved by some existing sophisticated algorithms, such as the
branch-and-bound method \cite{RLT}. For the second order terms in constraints (\ref{CONS3}) and
(\ref{CONS8}), we adopt a relaxation technique, the Reformulation-Linearization Technique (RLT)
\cite{mao, RLT} to linearize constraints (\ref{CONS3}) and (\ref{CONS8}). The RLT technique produces tight linear programming relaxations for an
underlying nonlinear and non-convex polynomial programming problem.
In the procedure, a variable substitution is applied for each nonlinear term to linearize the objective function
and the constraints of this problem. Besides, nonlinear implied constraints for each
substitution variable are generated by taking the products of
bounding terms of the decision variables, up to a suitable order. Specially, we present the RLT
procedures for (\ref{CONS3}) and (\ref{CONS8}) in the following.

For constraint (\ref{CONS3}), we define a substitution variable $s_{vpi}^k= {\delta ^k} \cdot
a_{vpi}^k$. Since the traffic of one link should be accommodated in one pairing, the number of time
slots of each pairing, $\delta ^k$, is bounded as $0\le{\delta ^k}\le{\widetilde{d}}$, where
${\widetilde{d}} = \max \{ \left\lceil {\frac{{{d_{vp}}}}{{{c_{vpi}}}}} \right\rceil
|{\rm{for}}\;{\rm{all}}\;v,p,i\} $. We also know $0\le{a_{vpi}^k} \le 1$. Then we can obtain the
\emph{RLT bound-factor product constraints} for $s_{vpi}^k$ as

\begin{equation}
\left\{ {\begin{aligned}&{\{ [{\delta ^k} - 0] \cdot [a_{vpi}^k - 0]\} _{LS}} \ge 0\\
&{\{ [\widetilde d - {\delta ^k}] \cdot [a_{vpi}^k - 0]\} _{LS}} \ge 0\\
&{\{ [{\delta ^k} - 0] \cdot [1 - a_{vpi}^k]\} _{LS}} \ge 0\\
&{\{ [\widetilde d - {\delta ^k}] \cdot [1 - a_{vpi}^k]\} _{LS}} \ge 0
\end{aligned}\;\;\;\;{\rm{for}}\;\;{\rm{all}}\;\;v,p,i,k}. \right.\\
\label{RLT CONS3-pre}
\end{equation}
${\{  \cdot \} _{LS}}$ represents a linearization step under $s_{vpi}^k= {\delta ^k} \cdot
a_{vpi}^k$. By substituting $s_{vpi}^k= {\delta ^k} \cdot
a_{vpi}^k$, we obtain

\begin{equation}
\left\{ {\begin{aligned}&{
{s_{vpi}^k \ge 0}}\\
&{\widetilde{d}  \cdot a_{vpi}^k - s_{vpi}^k \ge 0}\\
&{{\delta ^k} - s_{vpi}^k \ge 0}\\
&{ \widetilde{d}- {\delta ^k} - \widetilde{d}  \cdot a_{vpi}^k + s_{vpi}^k \ge 0 }
\end{aligned}\;{\rm{for}}\;\;\;{\rm{all}}\;\;v,p,i,k}. \right.\\
\label{RLT CONS3}
\end{equation}

For constraint (\ref{CONS8}), we first convert it to

\begin{equation}\hspace{-2.4cm}
\begin{aligned}
({k_0}{P_t}{l^{ - \gamma }_{s_{vpi},r_{vpi}}} \hspace{-0.1cm}-\hspace{-0.1cm} MS({c_{vpi}})W{N_0})\hspace{-0.1cm} \times \hspace{-0.1cm}
a_{vpi}^k \ge \\MS({c_{vpi}})\rho\sum\limits_{u = 1}^V {\sum\limits_{q = 1}^{{M_u}} {\sum\limits_{j = 1}^{{H_{uq}}}
{{f_{{s_{uqj}},{r_{vpi}}}}a_{vpi}^ka_{uqj}^k{k_0}{P_t}{l^{ - \gamma }_{{s_{uqj}},{r_{vpi}}}}} }}.\hspace{-2.4cm}
\end{aligned}
\end{equation}

For the second order term $a_{vpi}^ka_{uqj}^k$, we define $\omega_{vpi,uqj}^k= a_{vpi}^ka_{uqj}^k$ as the substitution
variable. Since $0 \le a_{vpi}^k \le 1$ and $0 \le a_{uqj}^k \le 1$, the \emph{RLT bound-factor product constraints} for $\omega_{vpi,uqj}^k$ are

\begin{equation}\hspace{-3cm}
\left\{ {\begin{aligned}&{
{\omega_{vpi,uqj}^k \ge 0}}\\
&{ a_{vpi}^k - \omega_{vpi,uqj}^k \ge 0}\\
&{a_{uqj}^k - \omega_{vpi,uqj}^k \ge 0}\\
&{1-a_{vpi}^k - a_{uqj}^k + \omega_{vpi,uqj}^k \ge 0 }
\end{aligned}}\right.\\\label{RLT bound-facotr constraints_2}
\end{equation}

After substituting $s_{vpi}^k$ and $\omega_{vpi,uqj}^k$ into constraint (\ref{CONS3}) and constraint (\ref{CONS8}), we obtain a mixed integer
linear program (MILP) relaxation (P2) as

\begin{equation}\hspace{-6.4cm}
 \min \sum\limits_{k = 1}^K
{{\delta ^k}} \label{OBJ_RF}
\end{equation}
\hspace{0.5cm}s. t.

\begin{equation}
\begin{array}{l}\hspace{-1cm}
\sum\limits_{k = 1}^K {s_{vpi}^k}\left\{ {\begin{array}{*{20}{c}}{ \ge \left\lceil
{\frac{{d_{vp}}}{{c_{vpi}}} } \right\rceil
,\;{\rm{if}}\;i \le {H_{vp}}},\\
{=0,{\rm{otherwise}};}\hspace{1cm}
\end{array}\;}\;\; \forall\;v,p,i\label{CONS3_RF} \right.\\
\end{array}
\end{equation}

\begin{equation}\hspace{0.4cm}
\begin{aligned}&
({k_0}{P_t}{l^{ - \gamma }_{s_{vpi},r_{vpi}}} \hspace{-0.1cm}-\hspace{-0.1cm} MS({c_{vpi}})W{N_0})\hspace{-0.1cm} \times \hspace{-0.1cm}
a_{vpi}^k \ge MS({c_{vpi}})\rho\\&   \sum\limits_{u = 1}^V {\sum\limits_{q = 1}^{{M_u}} {\sum\limits_{j = 1}^{{H_{uq}}}
{{f_{{s_{uqj}},{r_{vpi}}}}\omega_{vpi,uqj}^k{k_0}{P_t}{l^{ - \gamma }_{{s_{uqj}},{r_{vpi}}}}} }}. \hspace{0.3cm} \forall\;v,p,i,k
\end{aligned}
\end{equation}

\hspace{0.4cm}Constraints (\ref{CONS1}) and (\ref{CONS2}) with ``${\rm{if}}\;{d_{vp}}
> 0$'' removed;

\hspace{0.4cm}Constraints (\ref{CONS4}), (\ref{CONS5}), (\ref{CONS6}), (\ref{CONS7}),
(\ref{RLT CONS3}), and (\ref{RLT bound-facotr constraints_2}).\\

%



As an example, we consider the WPAN of six nodes showed in Fig. \ref{fig:mpmh operation}(b). For
the flow from $A$ to $B$ with 18 packets, we select three paths, path 1, path 2, and path 3, which
have been illustrated in Fig. \ref{fig:mpmh operation}(b). In this example, we assume for
any two nonadjacent links, $(v,p,i)$ and $(u,q,j)$, $f_{{s_{uqj}},\;{r_{vpi}}}$ is equal to 0. Then
we solve the problem of (P2) using the branch-and-bound method. The traffic distribution scheme is
to send 3 packets through path 1, 9 packets through path 2, and 6 packets through path 3. The
transmission phase of the schedule has 9 time slots, and is already illustrated in Fig.
\ref{fig:mpmh operation}(a). Compared with the single hop transmission scheme, MPMH reduces the number
of time slots for transmission by 50\%.

The formulated mixed integer linear program (MILP) problem (P2) is NP-hard. The number of
decision variables is $\mathcal{O}((V {P_{max}}  H_{max}  )^2K)$, and the number of constraints is
$\mathcal{O}(({V {P_{max}}  H_{max}})^2K)$, where $P_{max}$ is the maximum number of paths of
flows, and $H_{max}$ is the maximum number of hops of paths. Using the branch-and-bound algorithm
to solve this problem will take significantly long computation time, e.g., several minutes for a 5-node network in \cite{mao}, which is
unacceptable for practical mmWave WPANs where the duration of one time slot is only a few
microseconds \cite{mao}. Therefore, to implement MPMH scheduling in a practical mmWave WPAN,
heuristic algorithms with low computational complexity are needed.


\section{The MPMH Scheme}\label{S5}


To solve problem (P2), which boosts the spatial reuse via multiple paths multi-hop transmissions,
firstly we should select suitable transmission paths. Then since the traffic distributed on these
paths has a big impact on the efficiency of spatial reuse, traffic of each flow should be
distributed on its transmission paths efficiently. Finally, transmission scheduling is needed to
accommodate the traffic of all flows with a minimum number of time slots by fully exploiting
spatial reuse. Following the above ideas, in this section, we propose a multiple paths multi-hop
transmission scheme (MPMH), which consists of three heuristic algorithms respectively for the path
selection, traffic distribution, and transmission scheduling.

\subsection{Path Selection}\label{S5-a}

We set
the maximum number of hops for each selected path, and denote it by $H_{max}$. The path selection algorithm first finds out all the possible paths from the sender $s_v$ to the
receiver $r_v$ with the number of hops less than or equal to $H_{max}$, denoted by $P_c(v)$. To
exploit higher transmission ability of each path, each hop of these paths should have the same or
higher channel quality than the direct path of flow $v$. Besides, each path should have no loop.
Besides, if $c_v$ is equal to 0, i.e., the direct path of flow $v$ is blocked, flow $v$ will not be transmitted through its direct path.
Then the algorithm examines each path in $P_c(v)$ in order of non-increasing lowest transmission rate
on each path. The set of selected paths from $P_c(v)$ is denoted by $P_s(v)$. Since the lowest transmission rate on each path determines its transmission ability,
we should first select the paths with high transmission ability into $P_s(v)$. Besides, the paths
in $P_s(v)$ should have no common hop to avoid degradation of efficiency. To maximize spatial
reuse, the hops with the lowest transmission rate in $P_s(v)$ should be nonadjacent to enable
concurrent transmissions. Therefore, $|P_s(v)|$ is bounded by $\left\lfloor {n/2} \right\rfloor$
with a network of $n$ nodes.

We denote the set of flows that will be transmitted through multiple paths as $F_{mpmh}$, which is
selected according to (\ref{flow_requirement}). Then, the path selection algorithm should select
suitable paths for each flow $v$ in $F_{mpmh}$. We denote the sender of flow $v$ as $s_v$, and the receiver of flow $v$ as $r_v$. We
also denote the first node and the last node of path $p$ as $f_p$ and $l_p$ respectively. For link from $l_p$ to $i$, we denote its transmission rate by $c_{l_pi}$. The set of possible paths
from $s_v$ is denoted as $P(v)$, and $P(v)$ is initialized to the vertex of $s_v$. For
each path $p$, we denote the lowest transmission rate on it as $c_l(p)$, and the hop with the
lowest transmission rate as $h_l(p)$. We denote the set
of lowest transmission rate hops of paths in $P_s(v)$ is denoted as $H_l(P_s(v))$.

The pseudo-code of the path selection algorithm is presented in Algorithm \ref{alg:PS}. Lines 1--18
find out all the possible paths with the number of hops less than or equal to $H_{max}$, denoted by
$P_c(v)$. For each path in $P(v)$, the algorithm extends this path to generate new paths with no
loop, as indicated by lines 3--11. In line 5, the new hop extended from $p$ should have a higher or
same transmission rate than that of the direct path of flow $v$. After generating new paths from
$P(v)$, the old paths in $P(v)$ are removed as in line 10, and $P(v)$ is updated by the new set of
paths $P_{new}$, as in line 12. For each path $p$ in $P(v)$, if its last node is the receiver
$r_v$, then this path will be added to $P_c(v)$, and removed from $P(v)$, as indicated by lines
14--16. Lines 19--28 select the eventual set of transmission paths, $P_s(v)$, from $P_c(v)$. The
algorithm first obtains the path with the largest transmission ability, i.e., the lowest
transmission rate on each path, as indicated by line 20. Then if this path has no common hop with
the paths already in $P_s(v)$, and the hop on this path with the lowest transmission rate,
$h_l(p)$, is not adjacent to the hops in $H_l(P_s(v))$, this path will be selected into $P_s(v)$,
as indicated by lines 21--25. This path is removed from $P_c(v)$ in line 26. When there is no path
in $P_c(v)$, the algorithm outputs $P_s(v)$.

For the WPAN in Fig. \ref{fig:mpmh operation}(b), with $H_{max}$ set to 3, the path selection
algorithm obtains three paths for flow $A \to B$, i.e., path 1, path 2, and path 3 already
illustrated in Fig. \ref{fig:mpmh operation}(b).

\begin{algorithm}[htbp]
\caption{The Path Selection Algorithm} \label{alg:PS}
\begin{algorithmic}[1]
\REQUIRE ~~\\
 $P(v)=\{ {s_v}\} $; $P_c(v)=\emptyset$; $P_s(v)=\emptyset$;
 $H_l(P_s(v))=\emptyset$; $h$=0; \\
\ENSURE ~~\\
\WHILE {($|P(v)| > 0$ and $h<H_{max}$)}
\STATE  $h$=$h$+1;   $P_{new}=\emptyset$; \\
 \FOR {each $p\in P(v)$}
 \FOR {each node $i$ with link $l_p\to i$ unblocked}
\IF {($c_{{l_p}i}\ge c_v$ and $i$ is not on $p$)}
\STATE  Generate a new path $p^*$ by extending $p$ to $i$;\\
\STATE $P_{new}=P_{new}\cup p^*$;\\
\ENDIF \ENDFOR\\
\STATE $P(v)=P(v)-p$;\\
\ENDFOR\\
\STATE $P(v)=P_{new}$;\\
\FOR {each $p\in P(v)$} \IF {($l_p==r_v$)}
\STATE  $P_c(v)=P_c(v)\cup p$; $P(v)=P(v)-p$;\\
\ENDIF \ENDFOR\\
\ENDWHILE

\WHILE {($|P_c(v)| > 0$)} \STATE Obtain the path with the largest $c_l(p)$, $p\in P_c(v)$;
 \IF {($p$ have no common hop with paths in $P_s(v)$)}
 \IF {($h_l(p)$ is not adjacent to the hops in $H_l(P_s(v))$)}
\STATE  $P_s(v)=P_s(v)\cup p$; $H_l(P_s(v))=H_l(P_s(v)) \cup h_l(p)$;\\
\ENDIF \ENDIF \\
\STATE $P_c(v)=P_c(v)-p$;\\
\ENDWHILE \STATE Output $P_s(v)$.

\end{algorithmic}
\end{algorithm}

\subsection{Traffic Distribution} \label{S5-b}

With the results of path selection, we propose a traffic distribution algorithm to distribute the
traffic of a flow on the selected multiple paths, i.e., the traffic demand of flow $v$ should be
distributed on the set of selected paths, $P_s(v)$. Since the link with the lowest transmission
rate on a path determines the transmission ability of the path, we should let the links with lowest
transmission rates on different paths transmit concurrently as much as possible, and we also need
to maximize the utilization of time slots in a pairing by making as many links as possible to
transmit concurrently in each time slot of this pairing. Therefore, the traffic distribution
algorithm is the proportional distribution of the traffic according to the lowest transmission rate
on each path. For example, for the three paths in Fig. \ref{fig:mpmh operation}(b), since the
lowest transmission rates of path 1, path 2, and path 3 are 1, 3, and 2 respectively, the traffic
from A to B is distributed on path 1, path 2, and path 3 according to the proportion of 1:3:2.


\subsection{Transmission Scheduling} \label{S5-c}

After path selection and traffic distribution, the transmission scheduling algorithm computes a
near-optimal schedule to accommodate the traffic demand of flows. Since any two adjacent links cannot be scheduled concurrently, it can also be inferred that the
maximum number of links that can transmit in the same pairing is $\left\lfloor {n/2} \right\rfloor
$ \cite{mao}. Links that are scheduled in the same pairing can be represented by a directive graph, which is a matching
\cite{mao}. If we assign one of $K$ colors to each pairing, and all the edges in the same pairing have identical color, each hop will have only one of the $K$ colors since each hop is only scheduled in one pairing of the frame. Therefore, this process can also be modeled as an edge coloring problem \cite{mao}, and the only difference is that the hops on the same path should be scheduled one by one from the first hop to the last hop. We denote the set of directional links of the $t^{th}$ pairing as $H^t$, and the set of
vertices of the links in $H^t$ is denoted by $V^t$. Thus, the problem of optimal scheduling is to
find out the directive graph of each pairing to accommodate the traffic of all flows with a minimum
of time slots. Our algorithm starts scheduling from the paths with the largest number of hops, and
to maximize spatial reuse gain, the algorithm schedules as many links into each pairing as possible with
the condition of concurrent transmission satisfied. Since the inherent order of hops in the same
path, at most one hop can be scheduled in the same pairing, and the algorithm needs to schedule the
preceding hops first.

For each flow $v$, the set of its
transmission paths is denoted as $P_s(v)$, which can be obtained by the path selection algorithm.
For flows not in $F_{mpmh}$, its $P_s(v)$ only has the direct path. From the traffic distribution
algorithm, we obtains the traffic demand of each path in $P_s(v)$. In the transmission scheduling,
we denote the number of nodes as $n$, and the set of selected paths of all flows as $P_s$, which
includes the paths in $P_s(v)$ for each flow $v$. For each path $p\in P_s$, we denote its number of
hops as $h(p)$. For each hop of each path, according to the traffic distributed on this path, we
define the number of time slots to accommodate its traffic as the weight of this hop. For the
$i^{th}$ hop link of path $p\in P_s$, its weight is denoted as $w_{pi}$. We also denote the
$i^{th}$ hop link of path $p\in P_s$ as $(p,i)$. We denote the sender of link $(p,i)$ by $s_{pi}$, and the receiver by $r_{pi}$. The set of hops in $P_s$ is denoted by $H$. The
hop number of the first hop that is not scheduled on path $p$ is denoted as $F_u(p)$. In the
$t^{th}$ pairing, the set of paths that are not visited yet is denoted by $P_u^{t}$.

The pseudo-code of the transmission scheduling algorithm is presented in Algorithm \ref{alg:MPMH}.
First, we obtain the set of paths after path selection. From $P_s$, we can obtain the set of hops
in $P_s$, $H$. Since we should start scheduling from the first hop of each path, $F_u(p)$ for each
path $p$ is set to 1. Then we iteratively schedule each hop of $H$ into each pairing until all the
hops in $H$ are scheduled, as indicated in line 1. In each pairing, we first find out the unvisited
paths with the maximum number of unscheduled hops as in line 6. Then among these paths, the first
unscheduled hop with the minimum distance between its weight and current duration of this pairing
is selected as the candidate hop of this pairing, as indicated in line 7. This step makes the
numbers of time slots of hops in this pairing as close as possible, which can improve the
utilization of time slots in this pairing as much as possible. Then the algorithm checks whether
this candidate hop is adjacent to the hops already in this pairing since adjacent links cannot be
scheduled concurrently, as indicated by line 8. If it is not, first the candidate hop will be added
to this pairing, and then the concurrent transmission conditions of this pairing will be checked,
as indicated by lines 9--16. If the conditions cannot be met, this candidate hop will be removed
from this pairing, as indicated by lines 19--20. Otherwise, the duration of this pairing will
updated to accommodate the traffic of this hop as in line 17. Since at most one hop of one path can
be scheduled in one pairing, the path where this hop is will be removed from the set of the unvisit
paths as in line 22. When the number of links in each pairing reaches $\left\lfloor {n/2}
\right\rfloor $ or there is no path in the unvisited path set, the algorithm will start scheduling
for the next pairing as in line 5, and the scheduling for this pairing will be outputted as in line
24.

Applying this algorithm to the example in section \ref{S3-b}, we obtain the schedule as follows: in
the first pairing, link $A\to D$ of path 3 transmits for one time slots; in the second pairing,
link $A\to C$ of path 2 and $D \to F$ of path 3 transmit for three time slots; in the third
pairing, link $A\to B$ of path 1 and $C\to E$ of path 2 transmit for three time slots; in the
fourth pairing, link $F\to B$ of path 3 transmits for one time slot; in the fifth pairing, link
$E\to B$ transmits for two time slots. Thus, it needs 10 time slots to clear the traffic of $A\to
B$. As discussed before, the optimal solution needs 9 time slots. Thus our heuristic scheme obtains
nearly the same scheduling solution.

\begin{algorithm}[htbp]
\caption{The Transmission Scheduling Algorithm} \label{alg:MPMH}
\begin{algorithmic}[1]
\REQUIRE ~~\\
 Input: the set of selected paths of all flows, $P_s$;\\
 Obtain the set of hops in $P_s$, denoted by $H$; \\
 Obtain the number of hops for each path $p\in P_s$, $h(p)$; \\
 Obtain the weight of each hop $(p,i)\in H$, $w_{pi}$;\\
 Set $F_u(p)=1$ for each $p\in P_s$; $t$=0;
\ENSURE ~~\\
\WHILE {($|H| > 0$)}
\STATE  $t$=$t$+1;  \\
\STATE  Set ${V^t} = \emptyset $, ${H^t} = \emptyset $, and $\delta^{t}=0$; \\
\STATE Set $P_u^t$ with $P_u^t = P_s$; \\
\WHILE {($|{P_u^t}|>0$ and $|{H^t}| < \left\lfloor {n/2} \right\rfloor $)} \STATE Get the set of
unvisited paths with the largest
number of unscheduled hops, $P_{mh}$; \\
\STATE Get the hop $(p, F_u(p))$  of path $p\in{P_{mh}}$ with the minimum
${\rm{abs}}(\delta^{t}-w_{pF_u(p)})$;
 \IF {($s_{pF_u(p)} \notin {V^t}$\;and\;$ r_{pF_u(p)} \notin {V^t}$)}
\STATE  ${H^t} = {H^t} \cup \{ (p, F_u(p))\}$; \\\STATE ${V^t} = {V^t} \cup \{ s_{pF_u(p)},r_{pF_u(p)}\} $;\\
 \FOR {each link $(p,i)$ in ${H^t}$}
\STATE  Calculate the SINR of link $(p,i)$, $SIN{R_{pi}}$\\
\IF {($SIN{R_{pi}}<MS({c_{pi}})$)}
\STATE  Go to line 19\\
\ENDIF \ENDFOR
\STATE   $\delta^t={\rm{max}}(\delta^t, w_{pF_u(p)})$, $H=H-(p,F_u(p))$; \\ \STATE $F_u(p)=F_u(p)+1$; Go to line 21\\
\STATE   ${H^t} = {H^t} - \{ (p, F_u(p))\}$; \\ \STATE ${V^t} = {V^t} - \{ s_{pF_u(p)},r_{pF_u(p)}\}$;\\
\ENDIF\\
\STATE $P_u^t = P_u^t - p$;\\
\ENDWHILE\\

 \STATE Output $H^t$ and ${\delta ^t}$;\\
\ENDWHILE

\end{algorithmic}
\end{algorithm}


\subsection{Computational Complexity and Control Overhead} \label{S5-c}

The path selection algorithm has the computational complexity of $\mathcal{O}(n^{H_{max}})$, where
$n$ is the number of nodes in the network. For the traffic distribution algorithm, its
computational complexity is negligible. For the transmission scheduling algorithm, it has the
complexity of $\mathcal{O}(|P_s|n^2)$, where $P_s$ is the set of selected paths. Thus, the overall
complexity of our scheme is $\mathcal{O}(n^{H_{max}}+|P_s|n^2)$, which is a
pseudo-polynomial time solution and suitable for the implementation in practical mmWave WPANs.

With multi-path multi-hop transmissions, there will be increased control overhead since more scheduling information needs to be pushed to the nodes in the WPAN. However, the increase of the control overhead is small with respect to the Gbps transmission rates of mmWave links, and the influence on the performance is minimal.


\section{Performance Evaluation}\label{S6}
In this section, we evaluate the performance of our proposed MPMH under various traffic patterns,
and compare its performance with the optimal solution scheme and other existing protocols.

\subsection{Simulation Setup}

In the simulation, we consider a typical mmWave WPAN of 10 nodes. We assume that all the WNs and
the PNC are uniformly distributed in a square area with $8 m \times 8 m$. According to the
distances between WNs, we set four transmission rates, 2 Gbps, 4 Gbps, 6 Gbps, and 8 Gbps. There
are 10 flows in the network. With $\varepsilon $ of the criterion in Section \ref{S4-a} set to 0.0625, there is only one flow transmitted through multiple
paths, and other flows are transmitted through the direct paths. The paths are selected according to the algorithm in Section \ref{S5-a}. We set the size of
data packets to 1000 bytes. We adopt the simulation parameters in Table II of \cite{MRDMAC}, which is also summarized in Table \ref{tab:simulation_parameter}. The
duration of one time slot is set to 5 $\mu s$. With the transmission rate of 2 Gbps, a packet can
be transmitted in one time slot. As in \cite{mao}, for the simulated network, the PNC can complete
the polling of traffic or schedule pushing in one time slot. Generally, it takes only a few time
slots for the PNC to complete path selection and schedule computation. To adapt to dynamical
network states, the number of time slots of a frame is bounded by 1000. The simulation length is
set to $5 \times {10^4}$ time slots. We also set the delay threshold to $2.5 \times {10^4}$, and
will discard packets with delay larger than the threshold. Initially, each flow has a few packets
generated randomly. The maximum number of hops for each selected path, $H_{max}$, is set to 3. In the simulation, we assume nonadjacent links are able to be scheduled for concurrent transmissions.

\begin{table}
\begin{center}
\caption{Simulation Parameters}
\def \temptablewidth {0.9\textwidth}
\begin{tabular}{ccc}
\hline
\textbf{Parameter}&\textbf{Symbol}&\textbf{Value}\\
\hline
PHY data rate & R & 2 Gbps, 4 Gbps, 6 Gbps, 8 Gbps \\
Propagation delay&${\delta _p}$& 50ns\\
Slot Duration & $T_{slot}$ & 5 $\mu s$\\
PHY overhead& ${T_{PHY}}$ & 250ns\\
Short MAC frame Tx time& ${T_{ShFr}}$& ${T_{PHY}}$+14*8/R+${\delta _p}$\\
Packet transmission time&${T_{packet}}$& 1000*8/R\\
SIFS interval&${T_{SIFS}}$& 100ns\\
ACK Tx time&${T_{ACK}}$&${T_{ShFr}}$\\
\hline
\end{tabular}
\label{tab:simulation_parameter}
\end{center}
\end{table}


In the simulation, we set the following two kinds of traffic modes:

\subsubsection {\textbf{Poisson Process}} packets arrive at each flow following a poisson process with arrival
rate $\lambda $. The traffic load in this mode, denoted by ${T_l}$, is defined as
 \begin{equation}
{T_l} = \frac{{\lambda  \times L \times V}}{R}, \label{Tl_1}
\end{equation}
where $L$ is the size of data packets, $V$ is the number of flows, and $R$ is set to 2 Gbps.
\subsubsection {\textbf{Interrupted Poisson Process (IPP)}} packets arrive at each flow following an interrupted poisson
process (IPP). The parameters of the interrupted poisson process are ${{\lambda _1}}$, ${{\lambda
_2}}$, ${{p_1}}$ and ${{p_2}}$, and the arrival intervals of an IPP obey the second-order
hyper-exponential distribution with a mean of
\begin{equation}
E(X) = \frac{{{p_1}}}{{{\lambda _1}}} + \frac{{{p_2}}}{{{\lambda _2}}}.
\end{equation}

The interrupted Poisson process can be represented by an
ON-OFF process. The ON duration and the OFF duration obey the negative exponential
distribution of $r_1$ and $r_2$, respectively. In the ON duration, packets
arrive at each node following the Poisson process of arrival rate
${\lambda _{on\_off}}$, while no packet arrives in the OFF duration.
${\lambda _{on\_off}}$, $r_1$, and $r_2$ can be inferred from
${{\lambda _1}}$, ${{\lambda _2}}$, ${{p_1}}$, and ${{p_2}}$ as
follows.
\begin{equation}
{\lambda _{on\_off}} = {p_1}{\lambda _1} + {p_2}{\lambda _2},
\end{equation}
\begin{equation}
{r_1} = \frac{{{p_1}{p_2}{{({\lambda _1} - {\lambda
_2})}^2}}}{{{\lambda _{on\_off}}}},
\end{equation}
\begin{equation}
{r_2} = \frac{{{\lambda _1}{\lambda _2}}}{{{\lambda _{on\_off}}}}.
\end{equation}
Therefore, IPP traffic is typical bursty traffic. We define the traffic load ${T_l}$ in this mode as:
 \begin{equation}
{T_l} = \frac{{L \times V}}{{E(X) \times R}}.\label{Tl_2}
\end{equation}

We evaluate the system by the following four performance metrics:

1) \textbf{Average Transmission Delay:} The average transmission delay of received packets from all
flows, which is in units of time slots. The average transmission delay does not include the time caused by steering antenna.

2) \textbf{Network Throughput:} The number of successful transmissions of all flows until the end
of simulation. For a packet of one flow, if its delay is less than or equal to the threshold, it
will be counted as a successful transmission. With constant simulation length and fixed packet
size, total number of successful transmissions is a good indication to show the throughput
performance.

3) \textbf{Average Flow Delay:} The average transmission delay of the flow transmitted through
multiple paths.

4) \textbf{Flow Throughput:} The number of successful transmissions achieved by the flow
transmitted through multiple paths.

In order to show the advantages of MPMH, we compare MPMH with the following two protocols:

1) \emph{\textbf{FDMAC}}: The frame-based scheduling directional MAC protocol, and the core of
FDMAC is the greedy coloring (GC) algorithm, which can compute near-optimal schedules with respect
to the total transmission time with low complexity \cite{mao}. FDMAC is also a frame based
protocol, and in GC, the traffic demand of flows is accommodated by iteratively scheduling each
flow into each concurrent transmission pairing in non-increasing order according to traffic demand.
To the best of our knowledge, FDMAC achieves the best performance among the existing protocols,
such as MRDMAC \cite{MRDMAC} and MDMAC \cite{MDMAC}, by fully exploiting spatial reuse in mmWave
WPANs.

2) \emph{\textbf{FDMAC--UR}}: FDMAC with links' differences in the transmission rate not
considered, and the transmission rates of all links are set to 1 Gbps.

\subsection{Comparison with the Optimal Solution }

We first compare MPMH with the optimal solution of problem (P2) by the branch-and-bound method. Since obtaining the optimal
solutions takes extremely long time, we assume there is only one flow with the traffic to transmit,
and the traffic of this flow needs to be transmitted through multiple paths. The traffic load is
defined as in (\ref{Tl_1}) and (\ref{Tl_2}) with $V$ equal to 1. In this case, the delay threshold
is set to $3 \times {10^4}$.

\begin{figure}[htbp]
\begin{minipage}[t]{0.5\linewidth}
\centering
\includegraphics[width=1\columnwidth,height=1.15in]{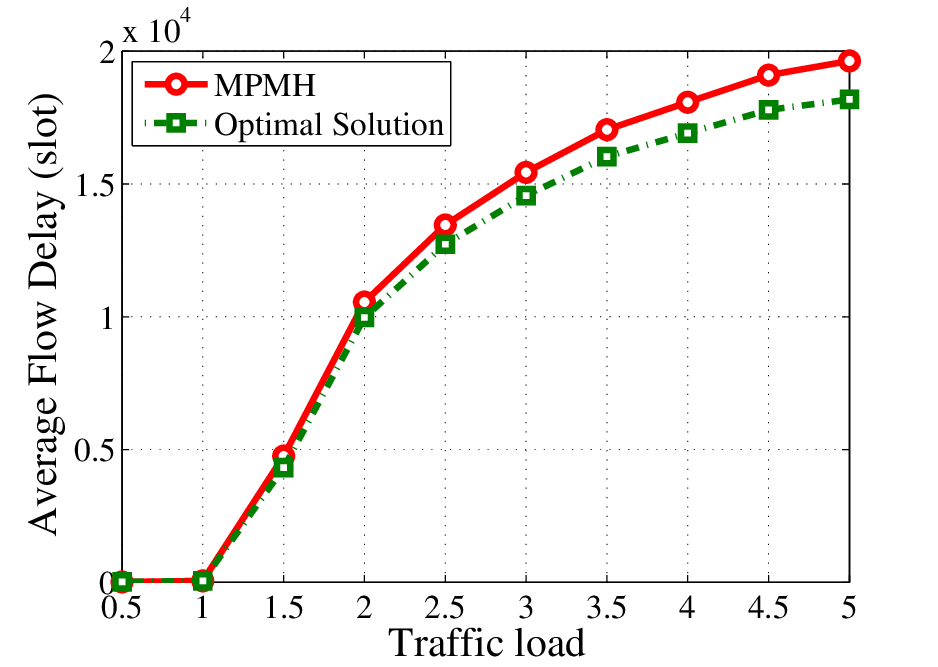}
\centerline{\small (a) Average flow delay}\label{FD:opt}
\end{minipage}%
\begin{minipage}[t]{0.5\linewidth}
\centering
\includegraphics[width=1\columnwidth,height=1.15in]{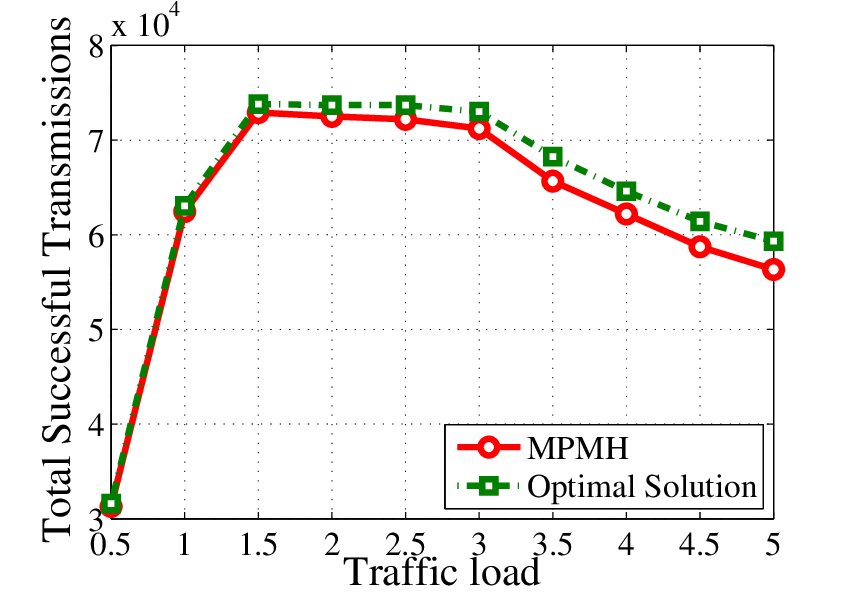}
\centerline{\small (b) Flow throughput}\label{FT:opt}
\end{minipage}%
\caption{Comparison between MPMH and the optimal solution under Poisson traffic.}
\label{fig:opt} 
\vspace*{-3mm}
\end{figure}

The average flow delay and the flow throughput of MPMH and the optimal solution are plotted in Fig.
\ref{fig:opt}. From the results, we can observe that the gap between MPMH and optimal solution is
negligible under light load. For the average flow delay, the gap at the traffic load of 5 is only
7.9\% of the average flow delay of MPMH. For the flow throughput, the gap is only 5.3\% of the flow
throughput of MPMH. Therefore, we have demonstrated that MPMH achieves near-optimal performance
with low complexity.

\begin{figure}[htbp]
\begin{minipage}[t]{1\linewidth}
\centering
\includegraphics[width=0.7\columnwidth]{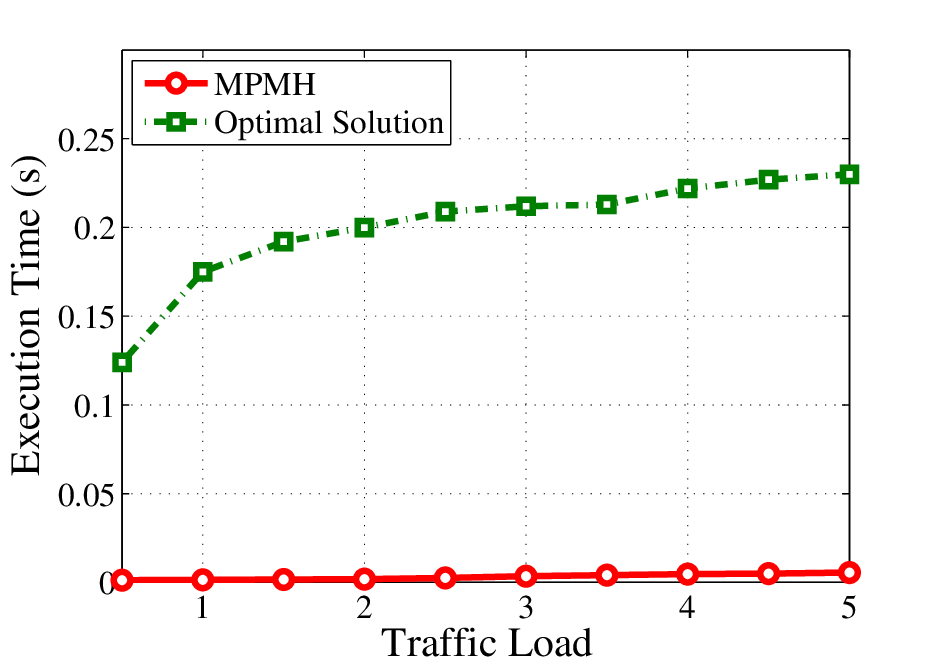}
\end{minipage}%
\caption{The execution time of MPMH and the optimal solution under Poisson traffic.}
\label{fig:time} 
\vspace*{-3mm}
\end{figure}

The average execution time of schedule computation for MPMH and the optimal solution is plotted in
Fig. \ref{fig:time}. From the results, we can observe that MPMH has much less execution time than
the optimal solution, and thus it has much lower computational complexity.

\subsection{Comparison with Existing Protocols}

\subsubsection{Delay}
\begin{figure}[htbp]
\begin{minipage}[t]{0.5\linewidth}
\centering
\includegraphics[width=1\columnwidth,height=1.15in]{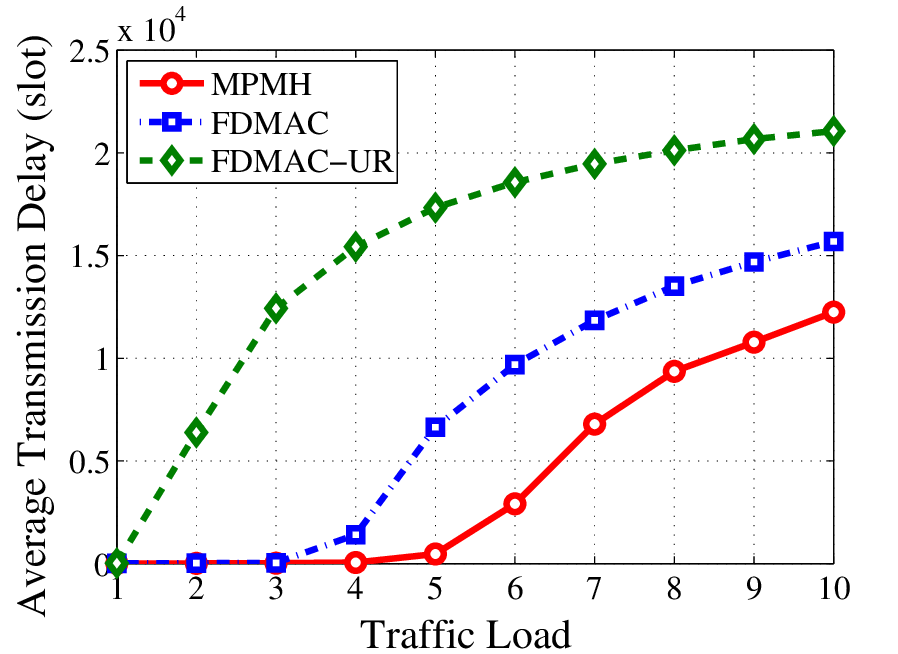}
\centerline{\small (a) Poisson traffic}
\end{minipage}%
\begin{minipage}[t]{0.5\linewidth}
\centering
\includegraphics[width=1\columnwidth,height=1.15in]{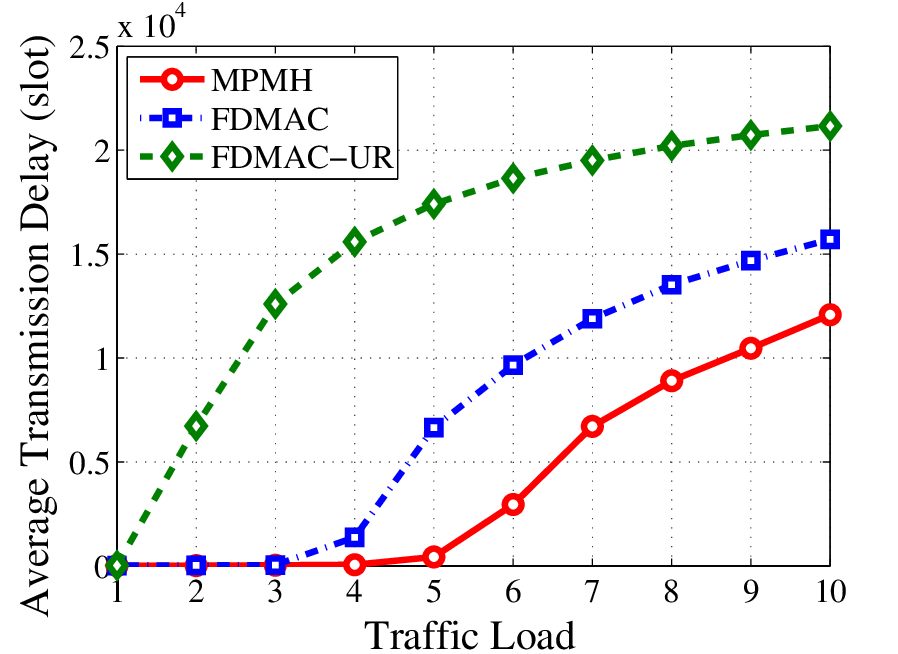}
\centerline{\small (b) IPP traffic}
\end{minipage}%
\caption{Average transmission delay of the three MAC protocols under different traffic loads.}
\label{fig:ND} 
\vspace*{-3mm}
\end{figure}

 We then evaluate the average network delay of the three MAC protocols.
 In Fig. \ref{fig:ND}, we plot it for different traffic
 loads. From the results, we can observe that with the increase of traffic load, the delay
 of the three protocols increases, and the delay of FDMAC-UR
 increases rapidly under light load. MPMH and FDMAC have similar delay performance under light
 load. Under light load, the arrived packets can be transmitted in a short time, and the frame length is short. Thus, the delay in this case is small. Under heavy load, MPMH outperforms FDMAC and FDMAC-UR. With the traffic load from 4 to 7, compared with FDMAC, MPMH decreases the
 average transmission delay by about 75.74\% under Poisson traffic and 86.54\%
 under IPP traffic on average. For FDMAC-UR, since the actual transmission
 ability of links is not exploited fully, it has much higher delay compared with MPMH and FDMAC.
 These phenomena can be explained as follows. In FDMAC and FDMAC-UR, flows cannot be transmitted through
 multiple paths of multiple hops. Thus, flows with low channel quality on their direct paths will occupy a large number of time slots in the
 schedule, which increases the frame length significantly and also delay of packets. Since the frame length is bounded by 1000 time slots, the packets that cannot be transmitted are re-scheduled in the next frame. Thus, with the increase of the traffic load, the system enters saturation, and the curves become flat and show a concave form.

\begin{figure}[htbp]
\begin{minipage}[t]{0.5\linewidth}
\centering
\includegraphics[width=1\columnwidth,height=1.15in]{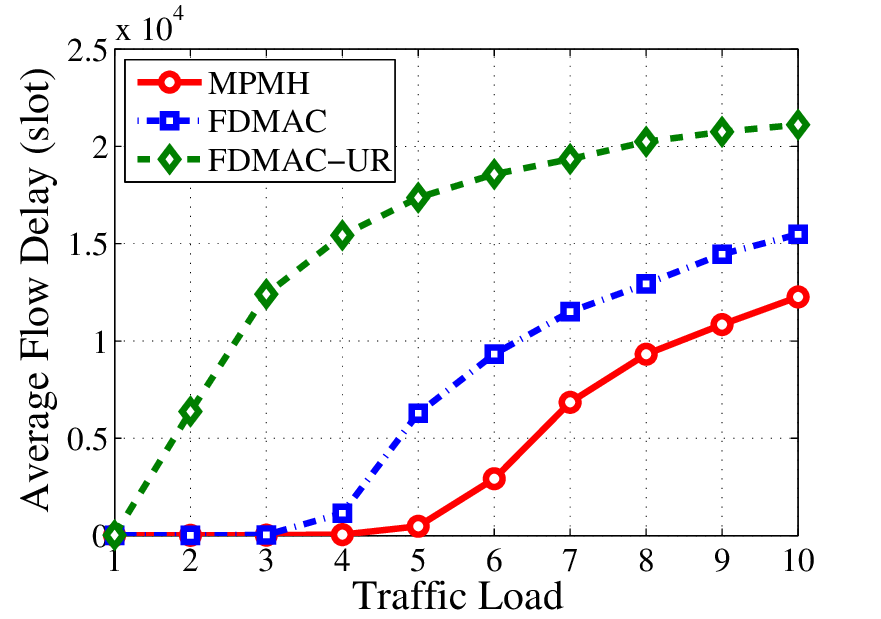}
\centerline{\small (a) Poisson traffic}
\end{minipage}%
\begin{minipage}[t]{0.5\linewidth}
\centering
\includegraphics[width=1\columnwidth,height=1.15in]{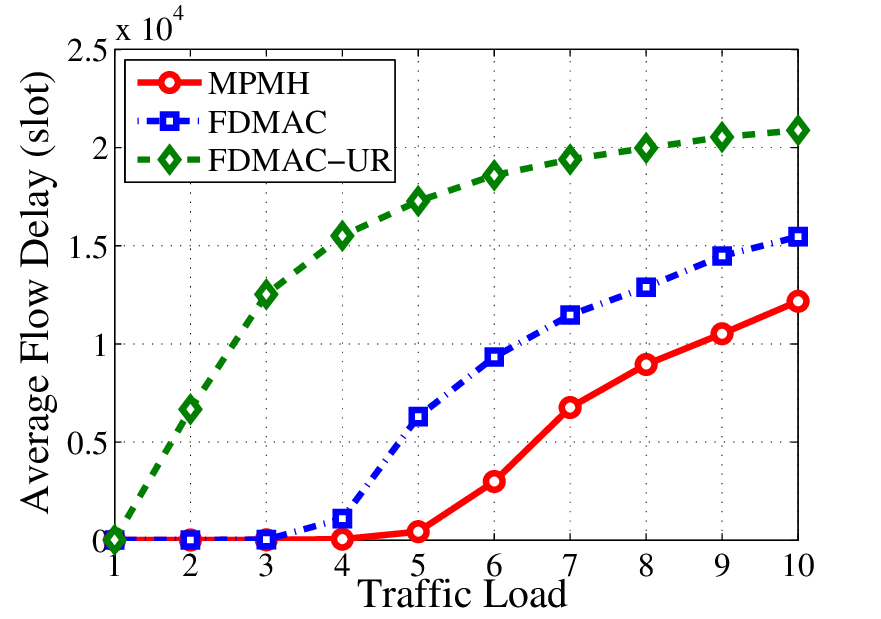}
\centerline{\small (b) IPP traffic}
\end{minipage}%
\caption{Average flow delay of the three MAC protocols under different traffic loads.}
\label{fig:FD} 
\vspace*{-3mm}
\end{figure}

 As regard to the average flow delay, we present the results in Fig. \ref{fig:FD}.
 We can observe that the delay curves of MPMH and FDMAC begin to
 diverge at the traffic load of 3. Compared with FDMAC, MPMH
 decreases the delay by about 74.31\% under Poisson traffic and
 74.29\% under IPP traffic on average with the traffic load from 4 to 7, respectively.

\subsubsection{Throughput}
\begin{figure}[htbp]
\begin{minipage}[t]{0.5\linewidth}
\centering
\includegraphics[width=1\columnwidth,height=1.15in]{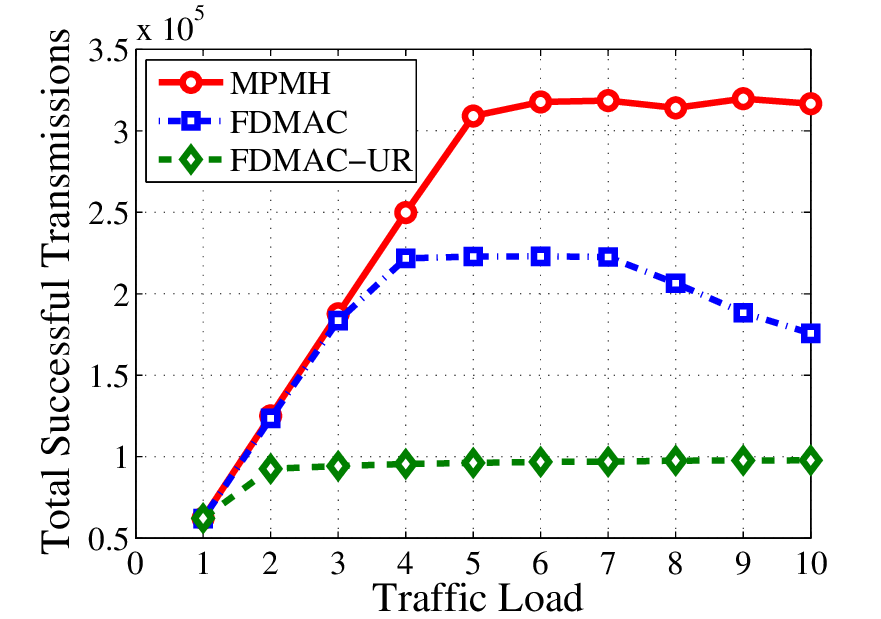}
\centerline{\small (a) Poisson traffic}
\end{minipage}%
\begin{minipage}[t]{0.5\linewidth}
\centering
\includegraphics[width=1\columnwidth,height=1.15in]{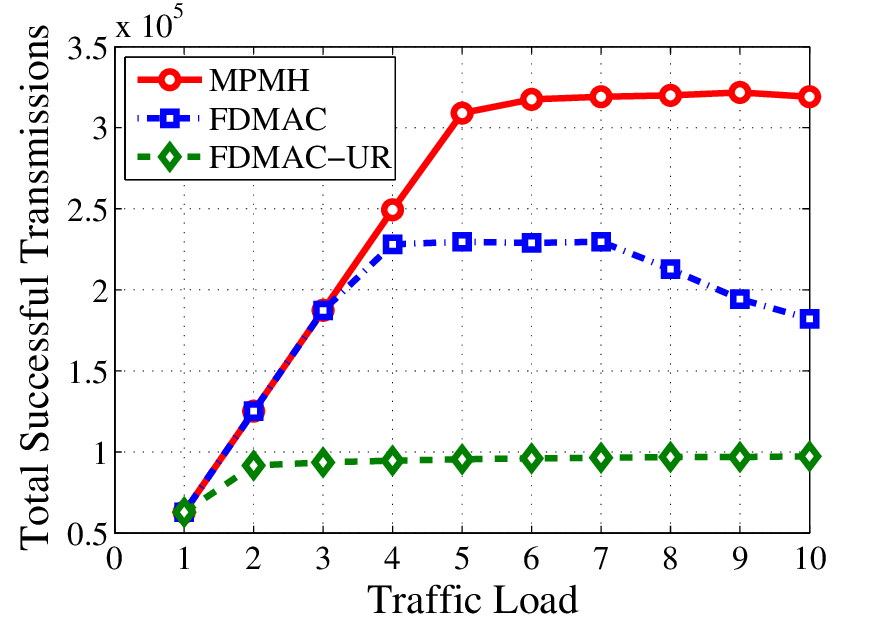}
\centerline{\small (b) IPP traffic}
\end{minipage}%
\caption{Network throughput of the three MAC protocols under different traffic loads.}
\label{fig:NT} 
\vspace*{-3mm}
\end{figure}

The network throughput achieved by the three protocols is plotted in Fig. \ref{fig:NT}. We can
observe that MPMH has the maximum throughput in all cases, and the gap between MPMH and FDMAC is
more significant under heavy load. Under light load, the delay is small, and all the arrived packets can be transmitted successfully for MPMH and FDMAC. Thus, the throughput of MPMH and FDMAC increase linearly under light load. FDMAC stops increasing at the traffic load of 4, while MPMH stops increasing at the traffic load of
6 since the network tends to saturation. For FDMAC-UR, its throughput is poor, and tends to be
saturated at the traffic load of 2. Compared with FDMAC, MPMH increases the network throughput with
the traffic load from 5 to 10 on average by about 54.37\% under Poisson traffic and about 50.58\%
under IPP traffic, respectively. When traffic load is 10, MPMH outperforms FDMAC by about 80.2\%.
The reason for MPMH to outperform FDMAC is that in FDMAC, flows with low channel quality on their direct paths cannot be
transmitted through multiple multi-hop paths as in MPMH, and occupy a large number of time slots in
the schedule, which are underutilized for concurrent transmissions. Under light traffic load, the
throughput of MPMH and FDMAC is determined by the traffic arriving at each node. Under heavy
traffic load, however, the advantages of MPMH over FDMAC will stand out, and MPMH will outperform
FDMAC significantly. On the other hand, with the increase of traffic load, delays of packets
increase, and a considerable number of packets are discarded due to their delays exceeding the
threshold, which leads to a decrease in FDMAC throughput from the traffic load of 7 and
the bigger gap between MPMH and FDMAC at the traffic load of 10.


\begin{figure}[htbp]
\begin{minipage}[t]{0.5\linewidth}
\centering
\includegraphics[width=1\columnwidth,height=1.15in]{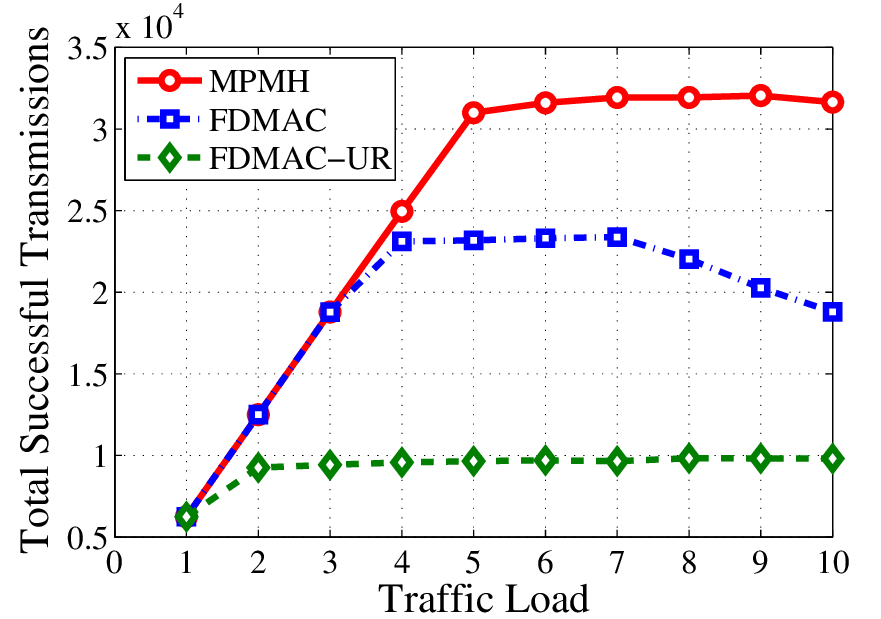}
\centerline{\small (a) Poisson traffic}
\end{minipage}%
\begin{minipage}[t]{0.5\linewidth}
\centering
\includegraphics[width=1\columnwidth,height=1.15in]{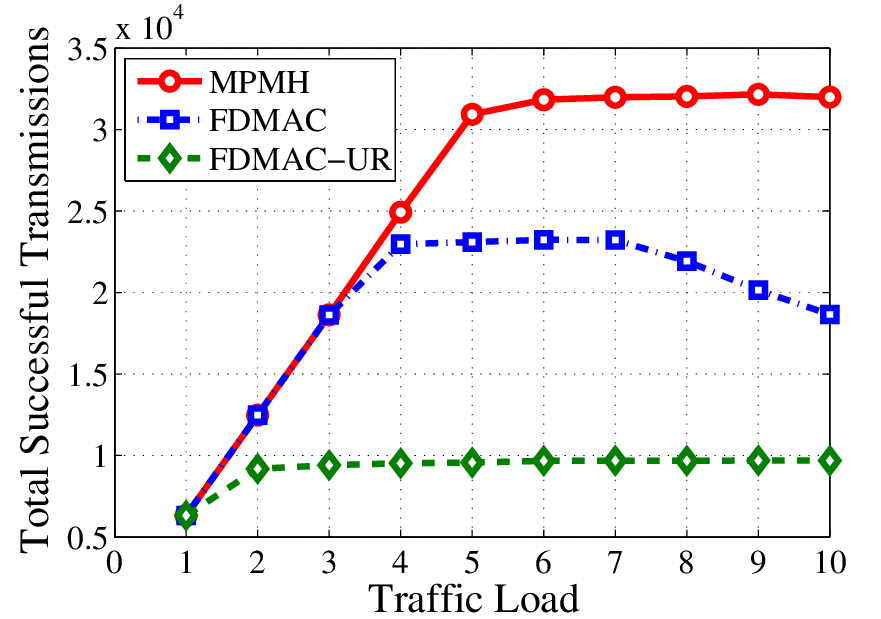}
\centerline{\small (b) IPP traffic}
\end{minipage}%
\caption{Flow throughput of the three MAC protocols under different traffic loads.}
\label{fig:FT} 
\vspace*{-3mm}
\end{figure}

On the other hand, the flow throughput of the three protocols is presented in Fig. \ref{fig:FT}. We
can observe that the tendencies of the curves are similar to those in Fig. \ref{fig:NT}. With the
traffic load from 5 to 10, MPMH outperforms FDMAC by about 52.14\% under Poisson traffic and
47.66\% under IPP traffic on average, respectively. The result show that MPMH greatly
increases the throughput of the flow of low channel quality compared to FDMAC, especially under
heavy traffic loads.

\subsection{Performance under different $H_{max}$}

\begin{figure}[htbp]
\begin{minipage}[t]{0.5\linewidth}
\centering
\includegraphics[width=1\columnwidth,height=1.15in]{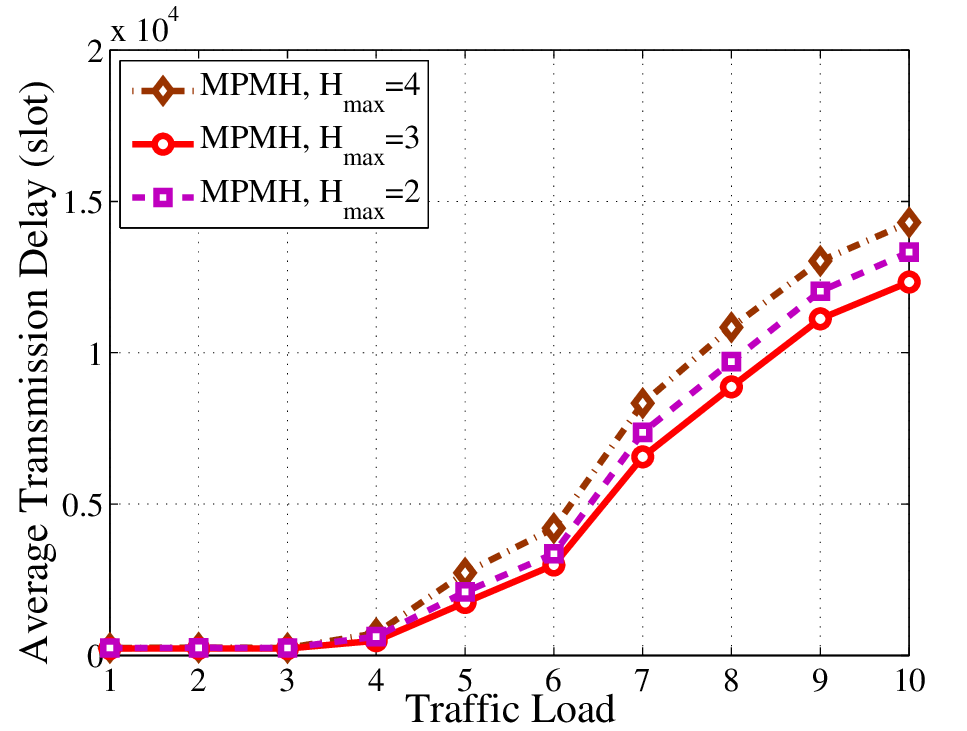}
\centerline{\small (a) Average Transmission Delay}
\end{minipage}%
\begin{minipage}[t]{0.5\linewidth}
\centering
\includegraphics[width=1\columnwidth,height=1.15in]{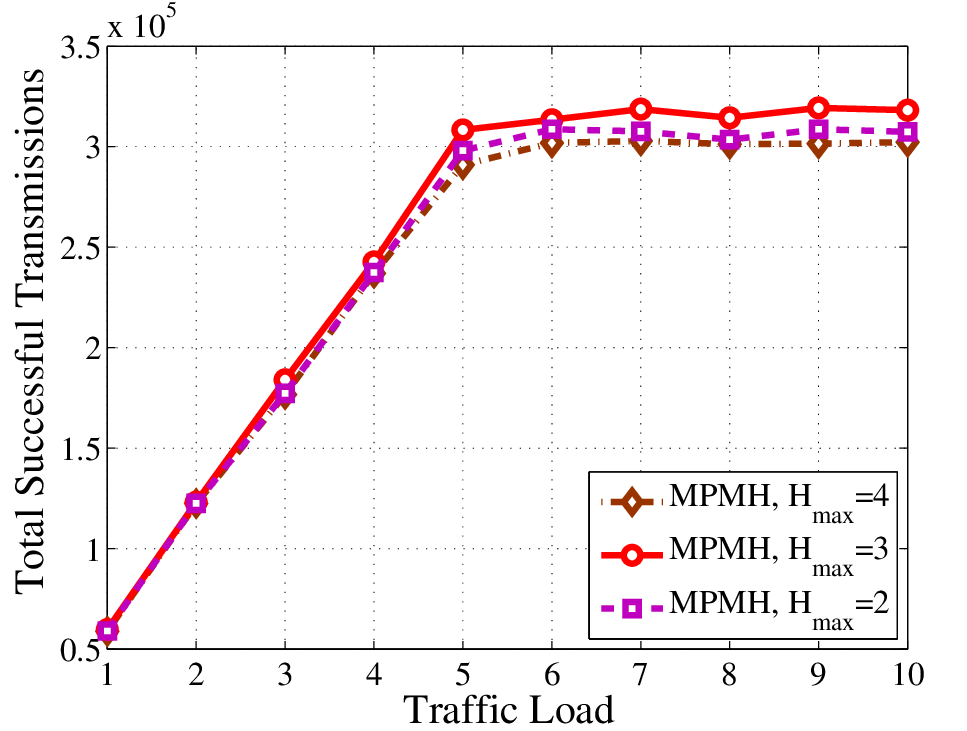}
\centerline{\small (b) Network Throughput}
\end{minipage}%
\caption{Delay and throughput of MPMH with different $H_{max}$ under Poisson traffic.}
\label{fig:H1}
\vspace*{-3mm}
\end{figure}

To investigate the impact of the choice of $H_{max}$ on network performance, we plot the average transmission delay and network throughput of MPMH with $H_{max}$ equal to 2, 3, and 4, respectively, in Fig. \ref{fig:H1}. The traffic mode is the Poisson traffic. From the results, we can observe that MPMH with $H_{max}$ equal to 3 achieves the best performance in terms of delay and throughput. However, the gap between MPMH with different $H_{max}$ is small. For MPMH with $H_{max}$ equal to 2, the advantage of multi-path multi-hop transmissions is limited by the number of hops on paths. For MPMH with $H_{max}$ equal to 4, more hops will lead to larger delay, and with the delays of more packets exceed the threshold, the network throughput also degrades. Therefore, the maximum number of hops on each path, $H_{max}$, should be optimized to achieve an optimal network performance in practice.

\begin{figure}[htbp]
\begin{minipage}[t]{0.5\linewidth}
\centering
\includegraphics[width=1\columnwidth,height=1.15in]{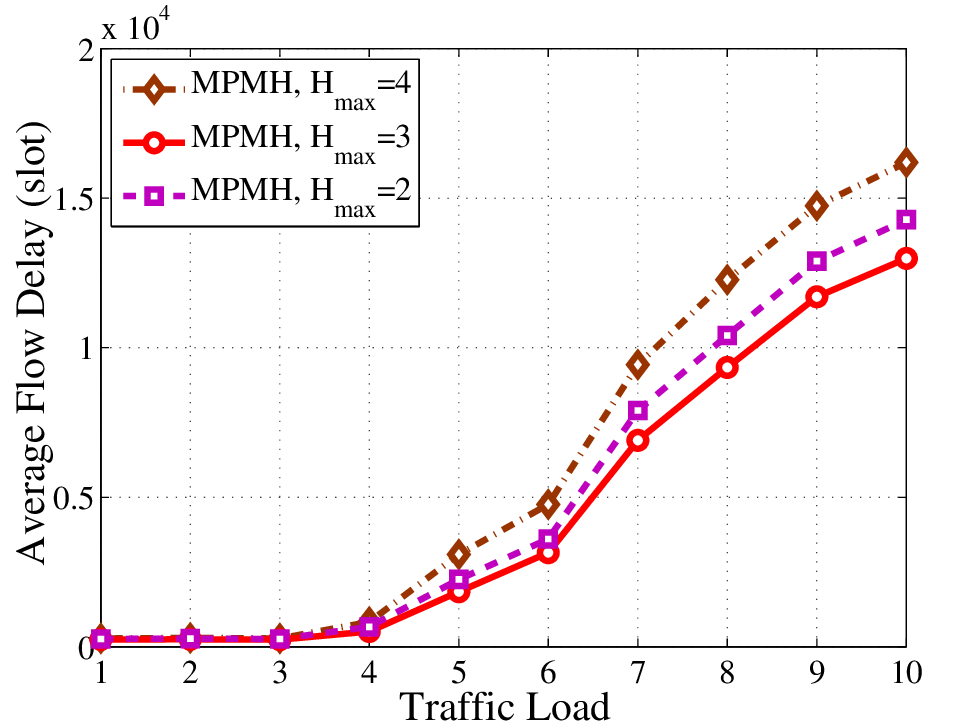}
\centerline{\small (a) Average Flow Delay}
\end{minipage}%
\begin{minipage}[t]{0.5\linewidth}
\centering
\includegraphics[width=1\columnwidth,height=1.15in]{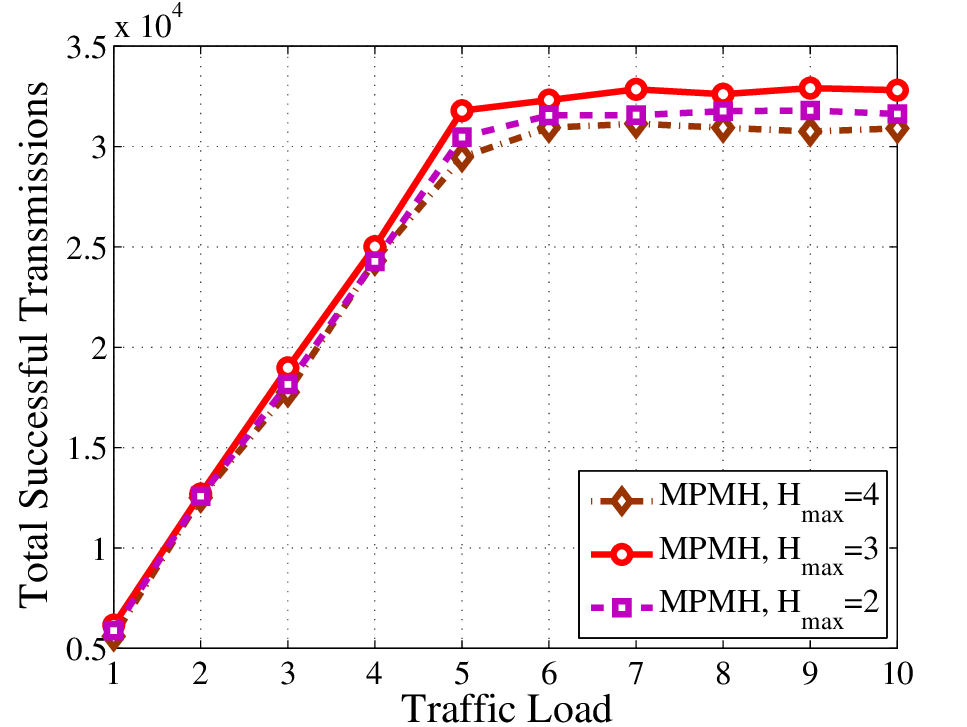}
\centerline{\small (b) Flow Throughput}
\end{minipage}%
\caption{Flow Delay and throughput of MPMH with different $H_{max}$ under Poisson traffic.}
\label{fig:H2}
\vspace*{-3mm}
\end{figure}

We also plot the flow delay and throughput of MPMH with $H_{max}$ equal to 2, 3, and 4, respectively, in Fig. \ref{fig:H2}. We can observe that the results are similar to those in Fig. \ref{fig:H1}, and MPMH with $H_{max}$ equal to 3 has the best performance in terms of flow delay and throughput. With a small $H_{max}$ as 2, the number of paths for each flow is limited, and the concurrent transmissions among paths cannot be exploited fully to improve flow delay and throughput performance. With a large $H_{max}$ as 4, transmissions through paths with more hops lead to larger delay, and there are more packets discarded since their delays exceed the threshold. In practice, $H_{max}$ should be selected according to the actual network conditions to optimize network performance.

\section{Conclusion}\label{S8}

In this paper, we proposed MPMH for mmWave WPANs in the 60 GHz band, which boosts the potential of
spatial reuse by transmitting the traffic of flows with low channel quality on their direct paths or with high traffic demand through multiple
multi-hop paths to improve throughput and to reduce latency for both these flows and the network.
Extensive simulations demonstrate that MPMH achieves near-optimal performance compared with the
optimal solution. Compared with FDMAC, MPMH increases flow and network throughput by about 50\% and
52.48\% on average, respectively. Performance under different maximum number of hops indicates the maximum number of hops should be selected according to the actual network conditions to optimize network performance.

In the future work, we will investigate the accurate and reasonable modeling of NLOS links and try to incorporate NLOS transmission into our scheme. In MPMH, the choice of $\varepsilon $ controls the number of flows transmitted through multiple paths, and we will further optimize $\varepsilon $ according to the actual network conditions such as the channel transmission rate distribution of links and the traffic demand distribution of flows. Although the fairness among flows is improved by the multi-path multi-hop transmission, we will investigate the mechanisms to improve the fairness of our scheme further. Besides, we will also evaluate the energy consumption and energy efficiency of MPMH.


    \begin{appendices}
      \section{Notation in Problem Formulation}
      In order to facilitate the reader to understand the notations in Problem Formulation, we list the notations in Table \ref{tab:notation1} as follows.

\begin{table}
\begin{center}
\caption{Notation in Problem Formulation}
\def \temptablewidth {0.9\textwidth}
\begin{tabular}{ll}
\hline
\textbf{Symbol}&\textbf{Description}\\
\hline
 $d_v$ & The traffic demand of flow $v$\\
 $M_v$ & The number of paths of flow $v$\\
 $H_{vp}$ & The number of hops of the $p^{th}$ path of flow $v$\\
 $d_{vp}$ & The traffic demand distributed to the $p^{th}$ path \\
 $(v,p,i)$ & The $i^{th}$ hop link of the $p^{th}$ path of flow $v$\\
 $c_{vpi}$ & The transmission rate of $(v,p,i)$\\
 \multirow{2}{*}{$I_{vpi,uqj}$} & A binary variable to indicate \\&if $(v,p,i)$ and $(u,q,j)$ are
adjacent\\
 $K$ & Number of pairings in a schedule \\
 $\delta^{k}$ & Number of time slots of the ${k^{th}}$ pairing \\
 \multirow{2}{*}{$a_{vpi}^k$} & A binary variable to indicate whether link $(v,p,i)$ \\&is scheduled in the ${k^{th}}$ pairing\\
 $s_{vpi}$ & Sender of link $(v,p,i)$\\
 $r_{vpi}$ & Receiver of link $(v,p,i)$\\
 \multirow{2}{*}{$f_{{s_{uqj}},\;{r_{vpi}}}$}& A binary variable to indicate whether $s_{uqj}$ and ${r_{vpi}}$ \\&direct their beams towards each other\\
 $l_{{s_{vpi}},\;{r_{vpi}}}$ & The distance between $s_{vpi}$ and $r_{vpi}$  \\
\hline
\end{tabular}
\label{tab:notation1}
\end{center}
\end{table}

      \section{Notation in MPMH}
      To facilitate the understanding of MPMH, we also list the notations in Table \ref{tab:notation2} as follows.

\begin{table}
\begin{center}
\caption{Notation in MPMH}
\def \temptablewidth {0.9\textwidth}
\begin{tabular}{ll}
\hline
\textbf{Symbol}&\textbf{Description}\\
\hline
 $H_{max}$ & The maximum number of hops for each selected path \\
 $s_v$ & The sender of flow $v$ \\
 $r_v$ & The receiver of flow $v$ \\
 $P(v)$ & Set of all possible paths from $s_v$\\
 $P_{new}$ & Set of generated new paths from $P(v)$ \\
 $P_c(v)$ & Set of all possible paths from $s_v$ to $r_v$\\
 $F_{mpmh}$ & Set of flows transmitted through multiple paths\\
 $f_p$ & The first node of path $p$\\
 $l_p$ & The last node of path $p$ \\
 $c_l(p)$ & The lowest transmission rate on path $p$\\
 $h_l(p)$ & The hop on path $p$ with the lowest transmission rate\\
 $P_s(v)$ & The set of selected paths\\
 $H_l(P_s(v))$ & The set of lowest transmission rate hops of paths in $P_s(v)$\\
 $\delta^{t}$ & Number of time slots of the ${t^{th}}$ pairing \\
 $H^t$ & Set of directional links in the $t^{th}$ pairing \\
 $V^t$ & Set of vertices of the links in $H^t$\\
 $n$ & The number of nodes \\
 $P_s$ & Set of selected paths of all flows \\
 $h(p)$ & Number of hops on $p$ \\
 $(p,i)$ & The $i^{th}$ hop link of path $p$ \\
 $s_{pi}$ & The sender of link $(p,i)$ \\
 $r_{pi}$ & The receiver of link $(p,i)$ \\
 $w_{pi}$ & The weight of link $(p,i)$\\
 $H$ & The set of hops in $P_s$\\
 $F_u(p)$ & The hop number of the first unscheduled hop on path $p$ \\
 $P_u^{t}$ & The set of unvisited paths\\
 \multirow{2}{*}{$P_{mh}$} & Set of unvisited paths with the largest \\&number of unscheduled hops\\
\hline
\end{tabular}
\label{tab:notation2}
\end{center}
\end{table}

  \end{appendices}

\bibliographystyle{IEEEtran}   

\begin{biography}[{\includegraphics[width=1in,height=1.25in,clip,keepaspectratio]{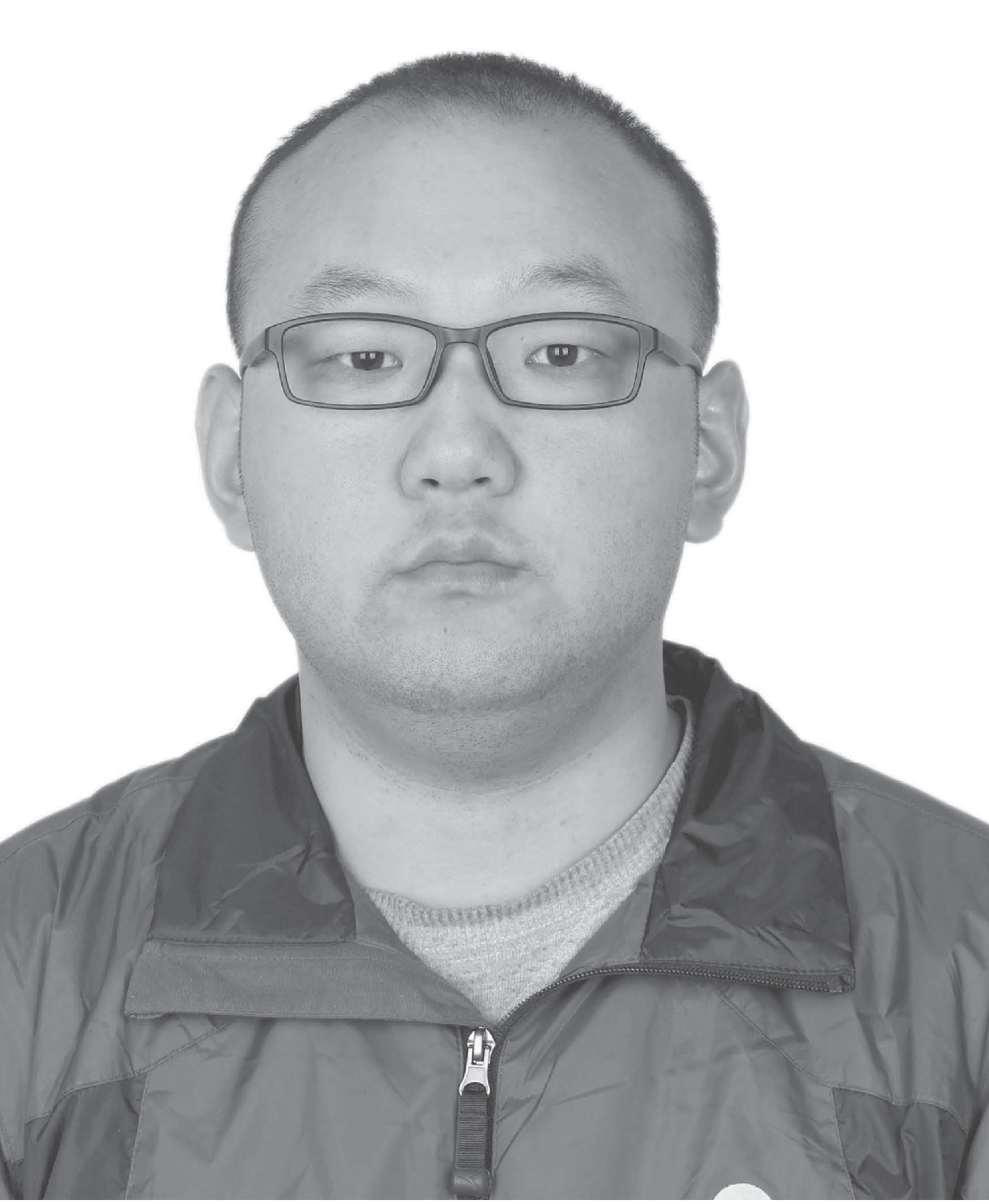}}]{Yong Niu}
received B.E. degree in Beijing Jiaotong University, China in 2011. He is currently working toward
his PhD degree at the Department of Electronic Engineering, Tsinghua University, China. His
research interests include millimeter wave communications, medium access control, and software-defined
networks.
\end{biography}

\begin{biography}[{\includegraphics[width=1in,height=1.25in,clip,keepaspectratio]{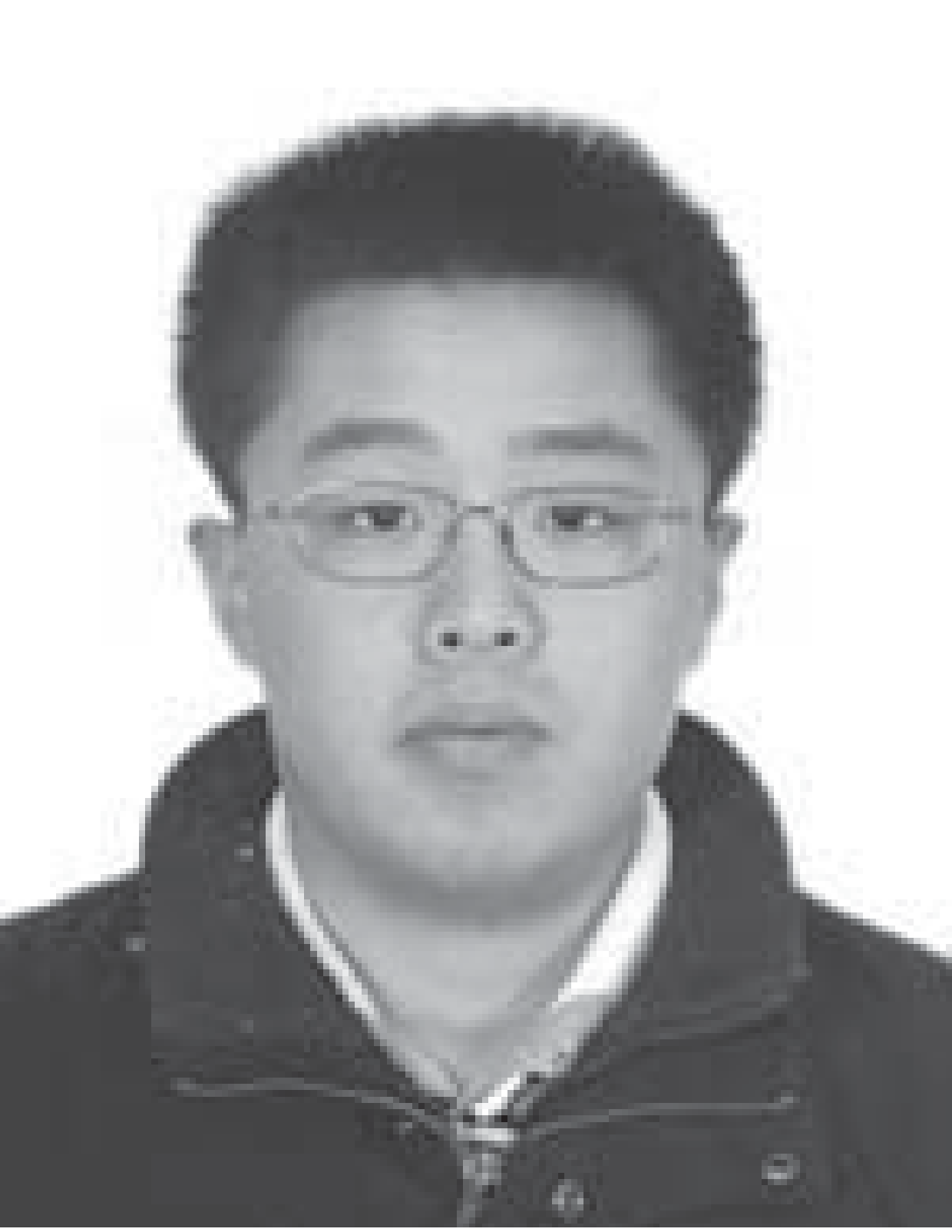}}]{Chuhan Gao}
is an undergraduate student from Tsinghua University, Beijing, China. He is currently working toward the B.E. degree. His research interests include millimeter-wave wireless communications, wireless networks, and software-defined networks.
\end{biography}

\begin{IEEEbiography}[{\includegraphics[width=0.9in,height=1.1in,clip,keepaspectratio]{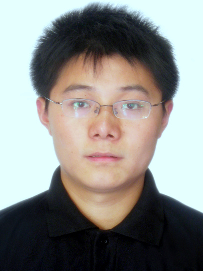}}]{Yong Li}
(M'2009) received his B.S. degree in Electronics and Information Engineering from Huazhong
University of Science and Technology, Wuhan, China, in 2007, and his Ph.D. degree in electronic
engineering from Tsinghua University, Beijing, China, in 2012. During July to August in 2012 and
2013, he worked as a Visiting Research Associate in Telekom Innovation Laboratories (T-labs) and HK
University of Science and Technology, respectively. During December 2013 to March 2014, he visited
University of Miami, FL, USA as a Visiting Scientist. He is currently a faculty member of the
Electronic Engineering at the Tsinghua University.\\ His research interests are in the areas of
networking and communications, including mobile opportunistic networks, device-to-device
communication, software-defined networks, network virtualization, future Internet, etc. He received
Outstanding Postdoctoral Researcher, Outstanding Ph. D Graduates and Outstanding Doctoral thesis of
Tsinghua University, and his research is granted by Young Scientist Fund of Natural Science
Foundation of China, Postdoctoral Special Find of China, and industry companies of Hitachi, ZET,
etc. He has published more than 100 research papers and has 10 granted and pending Chinese and
International patents. His Google Scholar Citation is about 440 with H-index of 11, also with more
than 120 total citations without self-citations in Web-of Science. He has served as Technical
Program Committee (TPC) Chair for WWW workshop of Simplex 2013, served as the TPC of several
international workshops and conferences. He is also a guest-editor for ACM/Springer Mobile Networks
\& Applications, Special Issue on Software-Defined and Virtualized Future Wireless Networks. Now,
he is the Associate Editor of EURASIP journal on wireless communications and networking.
\vspace*{-10mm}
\end{IEEEbiography}

\begin{biography}[{\includegraphics[width=1in,height=1.25in,clip,keepaspectratio]{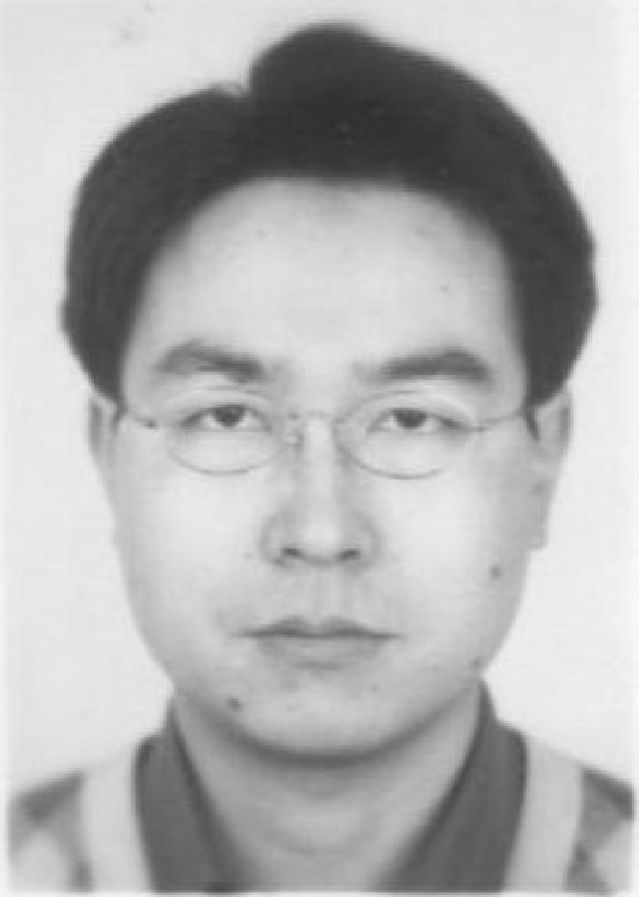}}]{Depeng Jin}
received the B.S. and Ph.D. degrees from Tsinghua University, Beijing,
China, in 1995 and 1999 respectively both in electronics engineering. \\
He is an associate professor at Tsinghua University and vice chair of Department of Electronic
Engineering. Dr. Jin was awarded National Scientific and Technological Innovation Prize (Second
Class) in 2002. His research fields include telecommunications, high-speed networks, ASIC design
and future Internet architecture.
\end{biography}

\begin{biography}[{\includegraphics[width=1in,height=1.25in,clip,keepaspectratio]{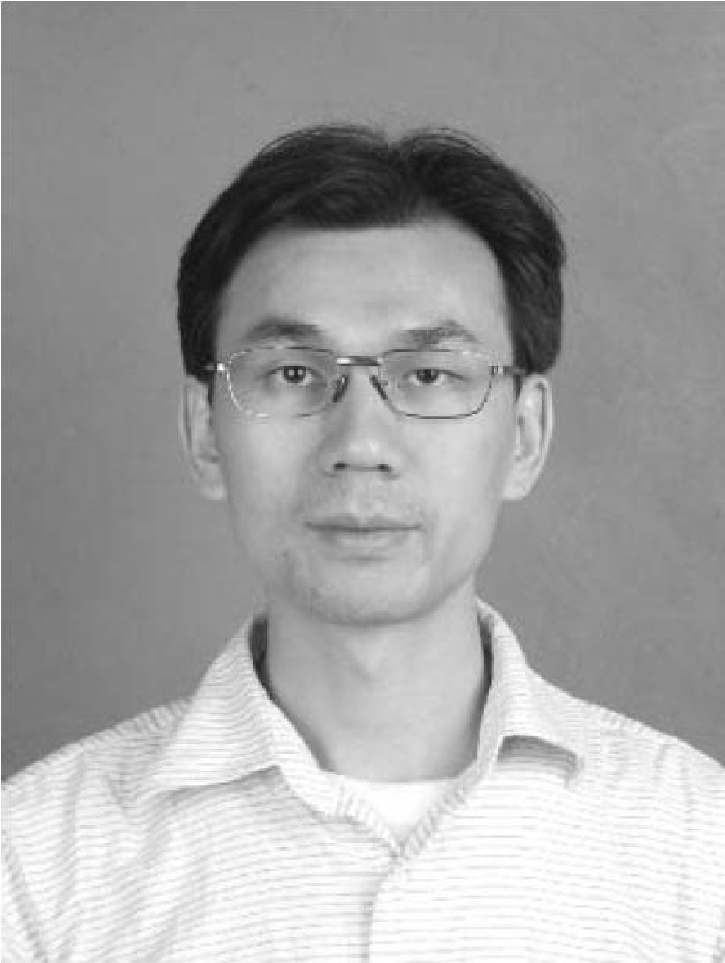}}]{Li Su} received the B.S. degree from Nankai University,
Tianjin, China, in 1999 and Ph.D. degree from Tsinghua University, Beijing, China in 2007
respectively both in electronics engineering. Now he is a research associate with Department of
Electronic Engineering, Tsinghua University. His research interests include telecommunications,
future internet architecture and on-chip network.
\end{biography}

\begin{IEEEbiography}[{\includegraphics[width=1in,height=1.25in,clip,keepaspectratio]{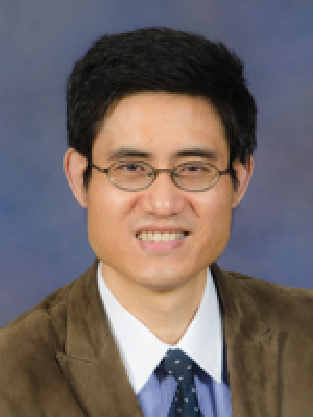}}]{Dapeng Oliver Wu}
Dapeng Oliver Wu (S'98--M'04--SM¡¯06--F'13)  received B.E. in
Electrical Engineering from Huazhong University of Science and
Technology, Wuhan, China, in 1990, M.E. in Electrical Engineering
from Beijing University of Posts and Telecommunications, Beijing,
China, in 1997, and Ph.D. in Electrical and Computer Engineering
from Carnegie Mellon University, Pittsburgh, PA, in 2003.

He is a professor at the Department of Electrical and Computer
Engineering, University of Florida, Gainesville, FL.  His research
interests are in the areas of networking, communications, signal
processing, computer vision, and machine learning. He received
University of Florida Research Foundation Professorship Award in
2009, AFOSR Young Investigator Program (YIP) Award in 2009, ONR
Young Investigator Program (YIP) Award in 2008, NSF CAREER award in
2007, the IEEE Circuits and Systems for Video Technology (CSVT)
Transactions Best Paper Award for Year 2001, and the Best Paper
Awards in IEEE GLOBECOM 2011 and International Conference on Quality
of Service in Heterogeneous Wired/Wireless Networks (QShine) 2006.

Currently, he serves as an Associate Editor for IEEE Transactions on
Circuits and Systems for Video Technology, Journal of Visual
Communication and Image Representation, and International Journal of
Ad Hoc and Ubiquitous Computing.  He is the founder of IEEE
Transactions on Network Science and Engineering.  He was the
founding Editor-in-Chief of Journal of Advances in Multimedia
between 2006 and 2008, and an Associate Editor for IEEE Transactions
on Wireless Communications and IEEE Transactions on Vehicular
Technology between 2004 and 2007. He is also a guest-editor for IEEE
Journal on Selected Areas in Communications (JSAC), Special Issue on
Cross-layer Optimized Wireless Multimedia Communications.  He has
served as Technical Program Committee (TPC) Chair for IEEE INFOCOM
2012, and TPC chair for IEEE International Conference on
Communications (ICC 2008), Signal Processing for Communications
Symposium, and as a member of executive committee and/or technical
program committee of over 80 conferences. He has served as Chair for
the Award Committee, and Chair of Mobile and wireless multimedia
Interest Group (MobIG), Technical Committee on Multimedia
Communications, IEEE Communications Society. He was a member of
Multimedia Signal Processing Technical Committee, IEEE Signal
Processing Society from Jan. 1, 2009 to Dec. 31, 2012.  He is an
IEEE Fellow.

\end{IEEEbiography}

\end{document}